\documentclass[twocolumn,tighten]{aastex7}

\usepackage{mathtools}
\usepackage{amssymb}
\usepackage{amsmath}
\usepackage{times}

\hypersetup{linkcolor=blue,citecolor=blue,filecolor=blue,urlcolor=blue}

\newcommand\aastex{AAS\TeX}

\def\target{{NGC\,3221}~}
\def\targetn{{NGC\,3221}}
\def\xmm{{\it XMM-Newton}~}
\def\chandra{{\it Chandra}~}
\def\xmmn{{\it XMM-Newton}}
\def\chandran{{\it Chandra}}

\def\apec{{\texttt{apec}}~}
\def\apecn{{\texttt{apec}}}
\def\pow{{\texttt{powerlaw}}~}
\def\pown{{\texttt{powerlaw}}}
\def\pex{{\texttt{pexrav}}~}

\def\ph{{\texttt{phabs}}~}

\def\zph{{\texttt{zphabs}}~}
\def\zphn{{\texttt{zphabs}}}

\def\bb{{\texttt{bbody}}~}

\def\diskbb{{\texttt{diskbb}}~}
\def\diskbbn{{\texttt{diskbb}}}

\def\hin{{{\rm H}\,{\sc i}}}
\def\hii{{{\rm H}\,{\sc ii}~}}
\def\ovii{{{\rm O}\,{\sc vii}~}}

\def\oviiin{{{\rm O}\,{\sc viii}}}
\shorttitle{\targetn: X-raying stellar disk in detail}
\shortauthors{Das \textit{et al.}}

\defcitealias{Luangtip2015}{L15}
\defcitealias{Asmus2014}{A14}
\defcitealias{Asmus2020}{A20}

\begin{document}

\title{Deep X-ray observation of \targetn: everything everywhere all at once}

\correspondingauthor{Sanskriti Das}
\email{snskriti@stanford.edu; dassanskriti@gmail.com}

\author[0000-0002-9069-7061]{Sanskriti Das}
\altaffiliation{Hubble Fellow}
\affil{Kavli Institute for Particle Astrophysics and Cosmology, Stanford University, 452 Lomita Mall, Stanford, CA\,94305, USA}
\email{dassanskriti@gmail.com}  
\author[0000-0002-4822-3559]{Smita Mathur}
\affil{Department of Astronomy, The Ohio State University, 140 West 18th Avenue, Columbus, OH 43210, USA}
\affil{Center for Cosmology and Astroparticle Physics, 191 West Woodruff Avenue, Columbus, OH 43210, USA}
\email{mathur.17@osu.edu}
\author[0000-0003-2192-3296]{Bret~D.~Lehmer}
\affiliation{Department of Physics, University of Arkansas, 226 Physics Building, 825 West Dickson Street, Fayetteville, AR 72701, USA}
\affiliation{Arkansas Center for Space and Planetary Sciences, University of Arkansas, 332 N. Arkansas Avenue, Fayetteville, AR 72701, USA}
\email{lehmer@uark.edu}
\author[0000-0003-0667-5941]{Steven W. Allen}
\affil{Kavli Institute for Particle Astrophysics and Cosmology, Stanford University, 452 Lomita Mall, Stanford, CA\,94305, USA}
\affil{Department of Physics, Stanford University, 382 via Pueblo Mall, Stanford, CA 94305, USA}
\affil{SLAC National Accelerator Laboratory, 2575 Sand Hill Road, Menlo Park, CA 94025, USA}
\email{swa@stanford.edu}
\author[0000-0001-6291-5239]{Yair Krongold}
\affiliation{Instituto de Astronomia, Universidad Nacional Autonoma de Mexico, 04510 Mexico City, Mexico}
\email{yair@astro.unam.mx}
\author[0000-0003-1880-1474]{Anjali Gupta}
\affil{Columbus State Community College, 550 E Spring St., Columbus, OH 43210, USA}
\email{agupta1@cscc.edu}

\begin{abstract}
We present a comprehensive analysis of 475\,ks (438\,ks unpublished \& 37\,ks archival) \xmmn/EPIC-pn observation of a nearby, highly inclined, star-forming, luminous infrared galaxy \target through spatial, temporal, and spectral information. We confirm the presence of a low-luminosity (presumably Compton-thick) AGN. The 0.4-12\,keV luminosity and the hardness ratio of the six ultra-luminous X-ray sources (ULX) previously identified in \chandra data exhibit diverse variability on day-scale. The collective emission from unresolved sources exhibits a different day-scale variability. We have also discovered two new predominantly soft ($<1$\,keV) sources. One of these has an enigmatic spectral shape featuring a soft component, which we interpret as a superbubble in \targetn, and a variable hard component from a compact object, unresolved from the superbubble. We do not confidently detect any X-ray emission from SN\,1961L. The hot gas in the ISM (out to $\pm$6\,kpc from the disk plane) and the extraplanar region (6--12\,kpc) both require two thermal phases at $\sim 0.15$\,keV and $\sim 0.55$\,keV. The $\sim 0.55$\,keV component is fainter in the ISM than the $\sim 0.15$\,keV component, but the emission from the latter falls off more steeply with disk height than the former. This makes the extraplanar region hotter and less dense than the ISM. The proximity of \target and the occurrence of the underluminous AGN offer a unique observing opportunity to study the hot diffuse medium along with nuclear and disk-wide point sources.  
\end{abstract}

\keywords{\uat{Luminous infrared galaxies}{946} --- \uat{Starburst galaxies}{1570} --- \uat{AGN host galaxies}{2017} --- \uat{Active Galactic Nuclei}{16} --- \uat{X-ray binary stars}{1811} -- \uat{Ultraluminous X-ray sources}{2164} --- \uat{Interstellar medium}{847} --- \uat{Interstellar plasma}{851} --- \uat{Hot ionized medium}{752} --- \uat{Superbubbles}{1656} --- \uat{High Energy astrophysics}{739} --- \uat{X-ray astronomy}{1810}} 

\section{Introduction}

In X-rays, the disks of galaxies are primarily comprised of two categories of sources: extended and compact objects. The extended objects consist of the diffuse hot interstellar medium (ISM) and supernova remnants/superbubbles, and the compact objects are classified as active galactic nuclei (AGN), X-ray binaries (XRB), and ultraluminous X-ray sources\footnote{Non-nuclear sources with $\rm L_X > 10^{39} erg\;s^{-1}$} (ULX) \citep{Persic2002,Gilfanov2022,Nardini2022,King2023,Kim2023}. XRBs offer valuable insights into the evolution of massive stars in binary systems and accretion processes onto stellar-mass objects. Due to their short ($\approx 10$\,Myr) evolutionary timescale, high-mass X-ray binaries (HMXBs) effectively trace recent star-formation, making them the dominant X-ray emitter in galaxies with high star formation rate \citep[SFR;][]{Shtykovskiy2007}. The X-ray luminosity function (XLF), i.e., the number of XRBs as a function of their X-ray luminosity, can be used to test binary evolution and population-synthesis models \citep{Lehmer2019}. The diffuse hot gas serves as an indicator of winds from massive stars and supernova-driven outflows, as its spatial distribution correlates with recent star formation sites located in spiral arms traced by H$\alpha$ emission \citep{Tyler2004}. The stellar feedback can eject some material outside the stellar disk, necessitating the simultaneous investigation of the hot ISM and the surrounding circumgalactic medium (CGM). AGNs provide information about accretion onto supermassive black holes, and powerful outflows that can potentially throw material into the CGM and even reach the intergalactic medium (IGM), thereby depriving the galaxy of star-forming fuel. In contrast to optical and infrared emissions, which can be confused with other gaseous and dusty components within the galaxy, X-ray emission provides an unambiguous way to identify an AGN. The variation in the AGN occurrence with star-forming activities \citep{Aird2019} suggests different timescales and mechanisms involved in the accretion, recycled inflows, and outflows in the baryon cycles within the stellar disk and between the disk and the CGM. 

The XRB population is usually studied in face-on (low-inclination) galaxies because that provides an unobstructed view of the stellar disk. In these galaxies, it is feasible to constrain the spatial distribution of hot diffuse gas \textit{in} the stellar disk, at the cost of an unknown scale height. Conversely, edge-on (high-inclination) galaxies inform us about the spatial extent of the hot gas \textit{around} the stellar disk. Thus, one can retrieve complementary information about the hot ISM and extraplanar region (also referred to as the disk-halo interface or the inner CGM) from galaxies of different inclinations. The AGN usually outshines the rest of the galaxy, enabling its detection in galaxies across inclinations, although the classification of an AGN might depend on its orientation with respect to the observer. 

\begin{figure}
    \includegraphics[width=\linewidth]{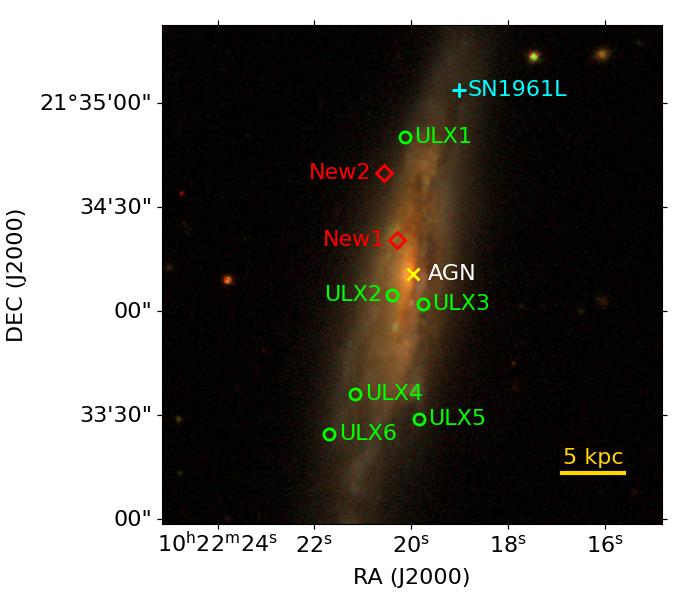}
    \caption{SDSS $gri$ (blue-green-red) image of \targetn. Positions of point sources (previously known or detected in this paper) are labeled. See \S\ref{sec:analysis} for more details.}
    \label{fig:optical}
\end{figure}

Observations of XRBs and the hot ISM in the same galaxy sample provide well-calibrated scaling relations between their respective X-ray luminosities and SFR \citep{Mineo2012a, Mineo2012b}, but studying the AGN along with other X-ray emitting components in the same galaxy is usually challenging due to the high luminosity of the AGN. Low-luminosity AGN (LLAGN) in star-forming galaxies in the nearby universe provide a unique observing condition, where the contribution from the AGN, XRB, and the surrounding hot gas can be separated and studied together in a system, providing a holistic view of an individual galaxy in the high-energy regime. That is the motivation of this paper.

In this paper, we present the new deep \xmmn/EPIC observations of a nearby highly-inclined star-forming galaxy \target (Figure\,\ref{fig:optical}) and provide a comprehensive picture of its diverse X-ray emitting constituents utilizing spatial, temporal, and spectral information. In \S\ref{sec:data}, we discuss the general properties of \target from the literature and introduce the X-ray data from various telescopes. In \S\ref{sec:reduction_xmm}, we outline the procedure for data reduction and extraction of images and spectra. In \S\ref{sec:analysis}, we analyze the data, show the results, and discuss their implications. Finally, in \S\ref{sec:summary}, we conclude with a summary of our findings with notes on potential directions for future studies. Throughout the paper, all errors have been quoted at 1$\sigma$.

\section{Existing data}\label{sec:data}

\target is an edge-on ($i=79^\circ$) SBcd luminous infrared \hii emission-line galaxy at $z=0.013379$ \citep{Samsonyan2016}. Its stellar mass is log(M$_\star$/M$_\odot$) = 11.00$\pm$0.10, calculated using B--V color and K-band luminosity, and the star formation rate (SFR) is 9.92$\pm$1.00 M$_\odot$ yr$^{-1}$, calculated using FUV flux and 8--1000 $\mu m$ total IR luminosity \citep{Lehmer2010}. Its high FIR surface brightness $\rm L_{FIR}/D_{25}^2 = 13.8 \times 10^{40}$ erg s$^{-1}$kpc$^{-2}$ and a high ratio of FIR flux densities, $S_{60\mu m}/S_{100\mu m} =0.37$ confirm it as an actively star-forming galaxy \citep{Rossa2003}. The detection of a supernova six decades ago \citep[SN\,1961L,][]{Barbon1999} also provides independent evidence of active star formation.

\target is not an AGN candidate based on its MIR color $W1 (3.4\mu m) - W2 (4.6 \mu m) = 0.185$ \citep[hereafter \citetalias{Asmus2020}]{Asmus2020}. There is no evidence supporting the presence of an AGN in the optical nebular emission lines, either. However, its W3 band ($12\mu m$) luminosity is quite high $10^{43.06}\rm erg\, s^{-1}$ similar to AGNs, although there could be a contribution from highly star-forming regions. \target satisfies the three-color demarcation criterion of AGN vs starburst of $0.17(W1-W2+24.5)>W2-W3$, putting it in the AGN category, unless the starburst s happening in regions with extremely high ionization parameters and/or column density \citep{Satyapal2018}. 

We observed \target (PI: Das) for 438\,ks as part of the \xmm large program AO-22 to constrain the density and temperature profiles of its hot circumgalactic medium (CGM), and assess the large-scale effect of stellar feedback on the CGM (Pan, Das \textit{et al.}, in prep.). Previous shallow \xmm data (37\,ks) have been used to study the images of its hot ISM and extraplanar region \citep{Tullmann2006} and the spectra of the integrated emission from the hot extended (30--200\,kpc) CGM \citep{Das2020a}. 

\target has also been observed with \chandra to study the point sources in its stellar disk \citep[\cite{Lehmer2010};][hereafter \citetalias{Luangtip2015}]{Luangtip2015} and the nucleus \citep{Torres-Alba2018}. It has $6_{-1}^{+0}$ point sources identified as ultraluminous X-ray binaries (ULX) and an additional hard X-ray source with $\Gamma = 0.3 \pm 0.6$ tentatively identified as an obscured AGN.

\section{Data reduction}\label{sec:reduction_xmm}
We reduce the \xmm data using a combination of Source Analysis Software (SAS v.21) and Extended Source Analysis Software (ESAS) tools. We have worked with unfiltered event files and reprocessed data using the up-to-date calibration database. We reduce both PN and MOS data and utilize them for point source detection. Because the effective area of the pn is larger than that of the MOS, we analyze only the pn data in the following steps to ensure a better S/N. Below, we outline the data reduction procedure: \\

1. We run \texttt{cifbuild} and \texttt{odfingest} to create the new CCF\footnote{current calibration file} Index File (CIF) and the observation directory file (ODF), respectively. 

2. We run the task \texttt{epchain} twice, once for the on-source and once for the OOT (out-of-time) exposure to process the event lists. 

3. We run \texttt{espfilt} to filter the data for soft proton flares. It creates a high-energy (2-12\,keV) count rate histogram from the data, fits the region around the peak of the histogram with a Gaussian, creates a good-time-interval (GTI) file for the time intervals with count rates within $\pm1.5\sigma$ of the best fitted peak, and filters the data within the GTI. We examine the count-rate histogram and the light curve to ensure no visually obvious residual contamination from the soft protons. The exposure times and GTI are provided in Table 1. 

\begin{table}[]
    \centering
    \caption{Details about observations}
    \begin{tabular}{cccc}
    \hline
    Instrument & obsID & texp[GTI] & Time \\
    & & [ks] & [DDMMYY] \\
    \hline
    XMM/EPIC & 0202730101 & 37.0[26.6] & 30/05/04 \\
    Chandra/ACIS-S & 10398 & 19.2 & 19/03/09 \\
    XMM/EPIC & 0922170101 & 108.6[72.2] & 03--04/05/23 \\
    XMM/EPIC & 0922170201 & 109.9[78.0] & 05--06/05/23 \\
    XMM/EPIC & 0922170301 & 109.6[79.6] & 10--12/05/23 \\
    XMM/EPIC & 0922170401 & 109.7[67.7] & 12--14/05/23 \\
\hline
    \end{tabular}  
    \label{tab:obsdetails}
\end{table} 

4. \xmm is not the ideal instrument to identify closely separated point sources and separate them from diffuse emission from the hot ISM. In our analysis, we use the sky positions of the point sources associated with \target from the \chandra data. However, we need to detect and remove the rest of the point sources in the FoV to assess the sky background correctly. We detect the point sources using \texttt{cheese} within 0.4--12\,keV. Beyond this energy range, the effective area of EPIC-pn is smaller, the energy calibration is uncertain, and the quiescent particle background (QPB) is higher. We tune the following parameters in \texttt{cheese} to optimize the source detection: 

$\bullet$ For the GTI of each obsID, we calculate the flux sensitivity of pn following \cite{Watson2001} for an $\alpha$= 1.7 powerlaw spectrum and Galactic N(\hin) $ = 1.76\times 10^{20}$ cm$^{-2}$ toward the direction of \target \citep{Bekhti2016}. We change the parameter \textit{rate}, the threshold of point-source flux,  from the default value 1.0 (in the unit of 10$^{-14}$ ergs cm$^{-2}$ s$^{-1}$) accordingly.

$\bullet$ The point source function (PSF) threshold parameter \textit{scale} is changed to 0.15 from the default value 0.25 to remove point sources down to a level where their flux is 15\% of its maximum value instead of 25\%. This allows us to remove a larger fraction of the point-source contamination. Reducing the value further did not improve the detectability of point sources. 

$\bullet$ Detection significance threshold \textit{mlmin} is reduced to 3 from the default value of 10 to ensure the inclusion of fainter sources in the filtered source list. 
   
$\bullet$ Minimum separation for point sources \textit{dist} is changed to 6$''$ from the default value 40$''$. This allows us to detect close-by sources. The on-axis PSF of EPIC-MOS is 6$''$ (FWHM); thus, we ensure that all the resolved sources are counted. \texttt{cheese} takes into account the spatial variation of off-axis PSF over the detector plane. 

5. The data are manually checked after the source detection. Because the disk of \target is X-ray bright with point sources and hot ISM, \texttt{cheese} identifies relatively brighter regions in the disk as point sources at arbitrary locations. We rewrite the \texttt{`emllist.fits'} file\footnote{the list of all the sources created by \texttt{emldetect} as part of the \texttt{edetect\_chain} task that is called by \texttt{cheese}} by excluding these sources and adding any visibly identifiable sources in the FOV that are not detected by \texttt{cheese}. Then we run \texttt{region} with the updated \texttt{`emllist.fits'} file to recreate the SAS and ESAS-compatible background region files in the sky and detector coordinates. Then we run \texttt{makemask} with the updated background region files to reconstruct the source removal mask. 

6. We run \texttt{pnspectra} to extract the spectra, science images, OOT images, and vignetting corrected exposure maps of the full FOV in supersoft (0.4--0.7\,keV), soft (0.7--1.0\,keV), medium (1--5\,keV), hard (5--12\,keV), and broad (0.4--12\,keV) bands. We run \texttt{pnback} to extract the QPB spectra and corresponding QPB images in those energy ranges and run \texttt{rotdet2sky} to construct those QPB images in sky coordinates.

7. For each point source, we define the ``on-source" region as a circle of 15$''$ diameter ($\approx$half-energy width) centered at the position of each source. We define the corresponding background region as a circular annulus of inner and outer diameters of 30$''$ and 60$''$ centered at the respective source. Because both the ``on-source" and background regions include diffuse ISM emission, choosing the background region this way ensures negligible residual ISM emission in the background-subtracted counts. We additionally exclude circular regions of 30$''$ diameter around all point sources to mitigate surrounding point source contamination. Examples of the resulting regions are shown in Figure\,\ref{fig:agn} and \ref{fig:new}.

8. For the ``total diffuse medium" (ISM + extraplanar region), we define the source as an ellipse that encompasses the outermost contour of the X-ray-bright disk visible in the images created in the previous step. The ellipse is centered at the optical/IR center of \target (RA, DEC = $10^h22^m20.07^s,+21^d34^m02.34^s$) and aligned with its stellar disk (PA=165$^\circ$), with semi-major and semi-minor axes of 55$''$ and 40$''$. We define the extraplanar region using the same outer boundary as the ``total diffuse medium" and exclude an inner ellipse with semi-major and semi-minor axes of 55$''$ and 20$''$, respectively. The background for these extended media is an elliptical annulus with outer semi-major and semi-minor axes of $\sqrt2\times 55''$ and $\sqrt2\times 40''$, and inner semi-major and semi-minor axes of $55''$ and $40''$\footnote{Because the diffuse medium could be more extended beyond the boundary of the contours, defining the background immediately outside the ellipse might overestimate the sky background. Thus, we obtain a conservative estimate of the ISM and extraplanar region. The extended CGM will be explored in a follow-up paper (Pan, Das \textit{et al.}, in prep.).}. Same as the previous step, we additionally exclude circular regions of 30$''$ diameter around all point sources to mitigate point source contamination. Examples of the resulting regions are shown in Figure\,\ref{fig:ism} (top row).

9. For point sources with sufficient count (see \S\ref{sec:spectroscopy}), we run \texttt{especget} to obtain source spectra, the RMF, ARF (corrected for encircled energy fraction), and the sky background spectra corresponding to the background region defined above. We run \texttt{grppha} to group channels and ensure at least 16 counts in each spectral bin. 

10. For the diffuse medium, we run the same \texttt{especget} command with the on-source and background regions as defined above, with the condition of \texttt{extendedsource} turned on to correctly calculate the ARF by averaging the mirror vignetting over the whole on-source region and not convolving the area lost due to bad pixels and chip gaps with the PSF. Then we group channels requiring at least 50 counts in each spectral bin by running \texttt{grppha}. 




\section{Data analysis, results, and discussion}\label{sec:analysis}

\subsection{Imaging}\label{sec:imaging}

\begin{figure*}
    \centering
    \includegraphics[trim=0 0 0 0, clip,width=0.195\linewidth]{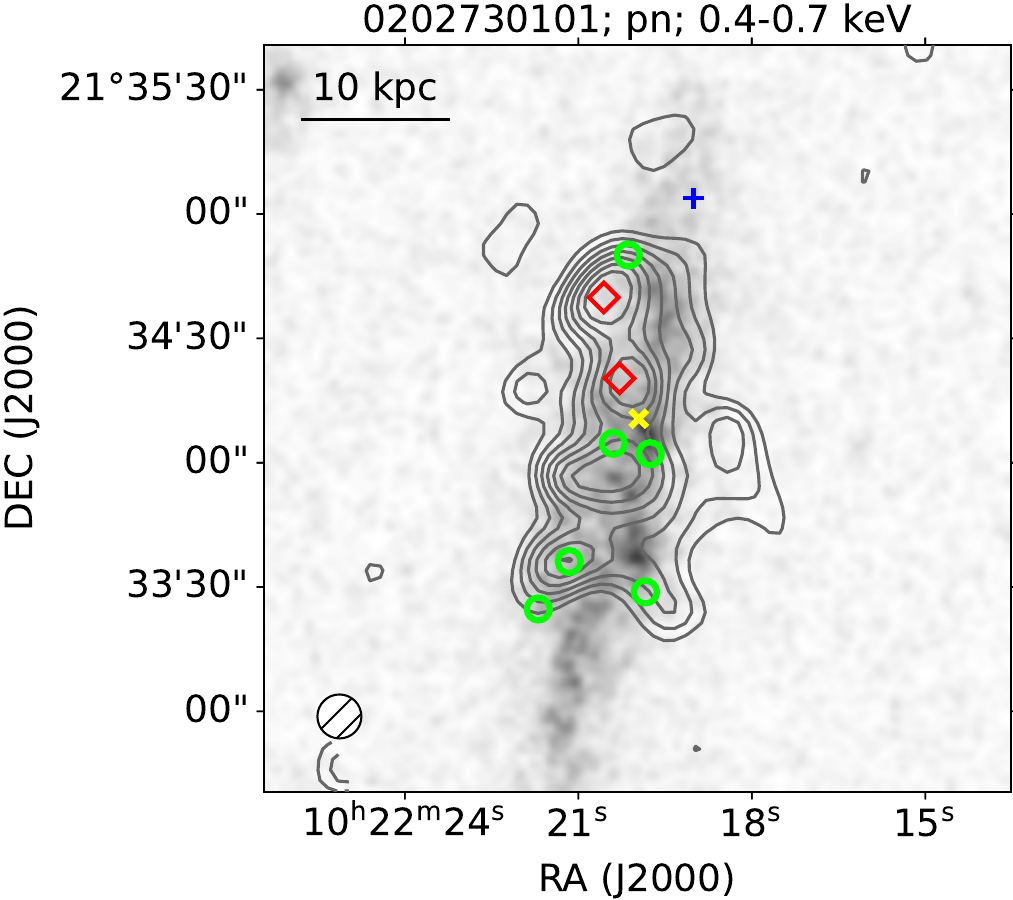}
    \includegraphics[trim=0 0 0 0, clip,width=0.195\linewidth]{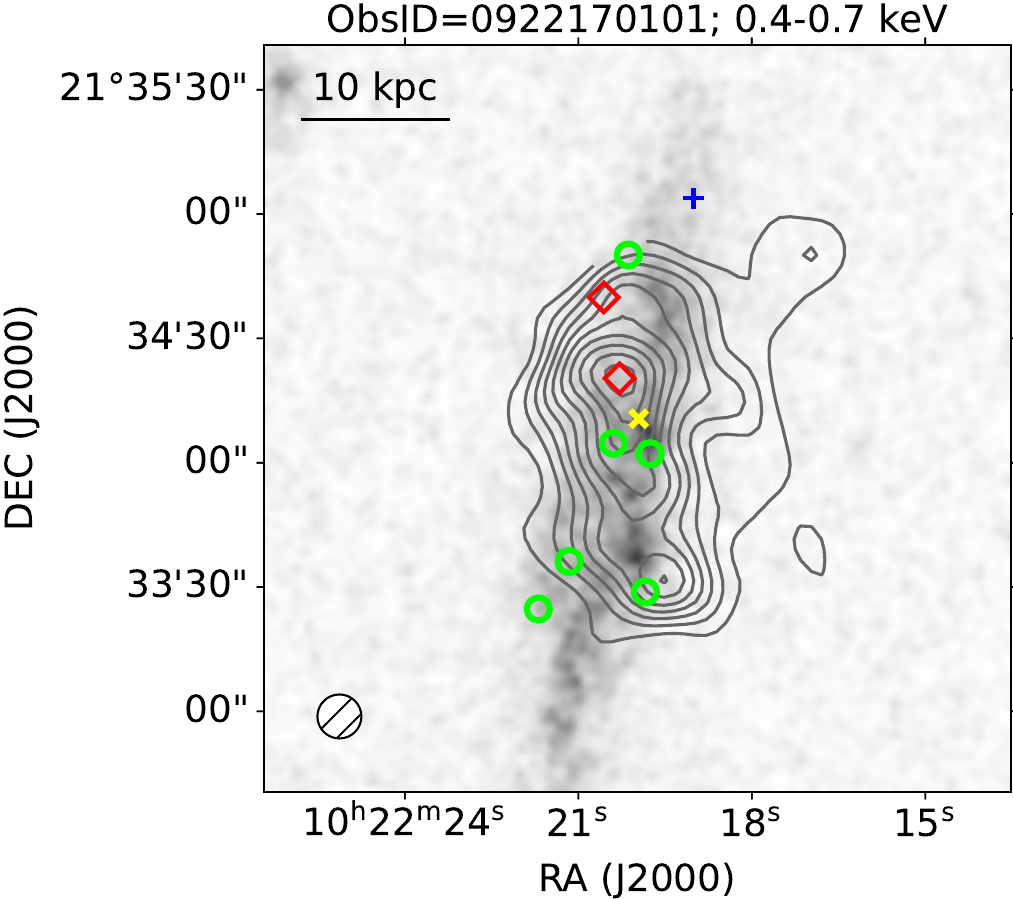}
    \includegraphics[trim=0 0 0 0, clip,width=0.195\linewidth]{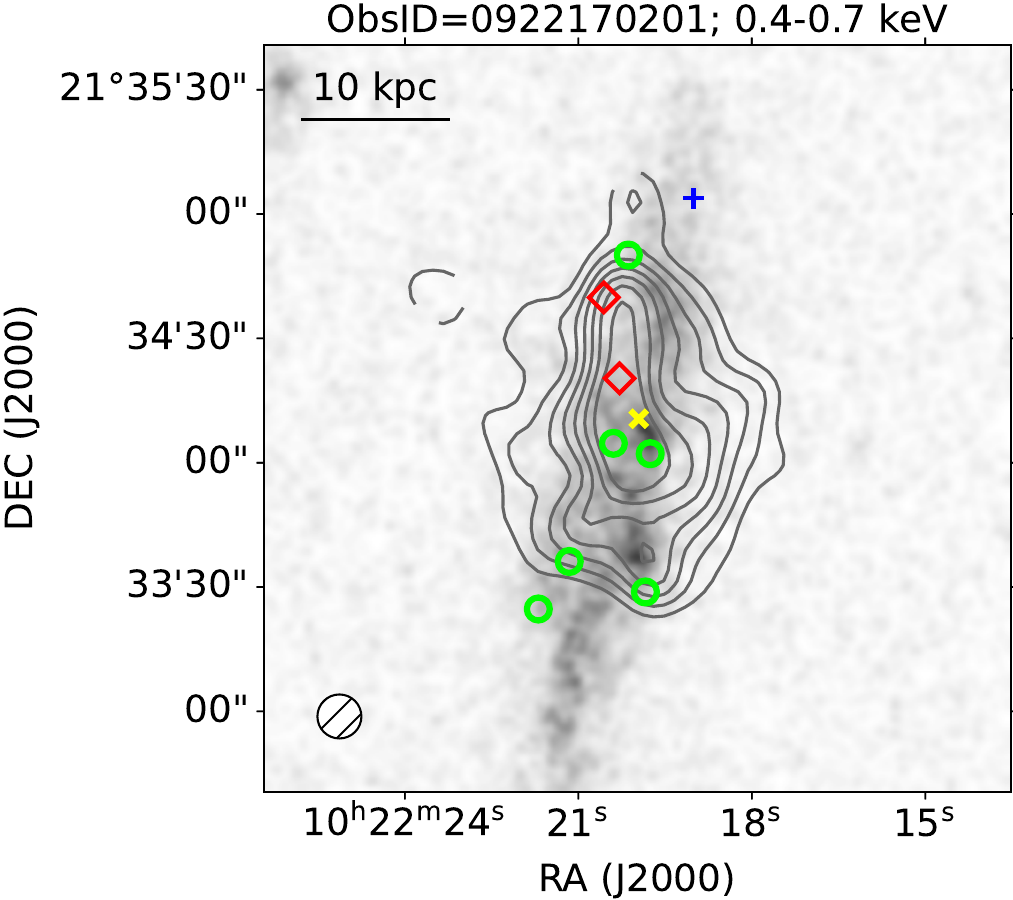}
    \includegraphics[trim=0 0 0 0, clip,width=0.195\linewidth]{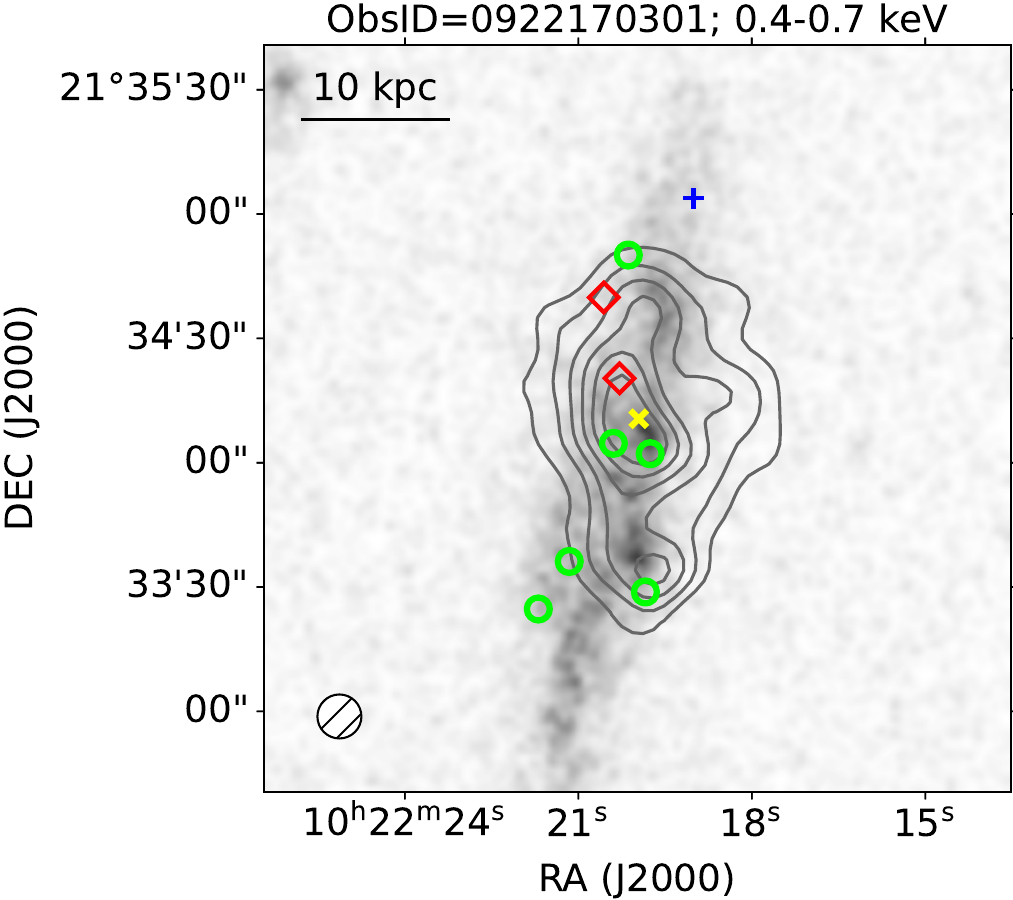}
    \includegraphics[trim=0 0 0 0, clip,width=0.195\linewidth]{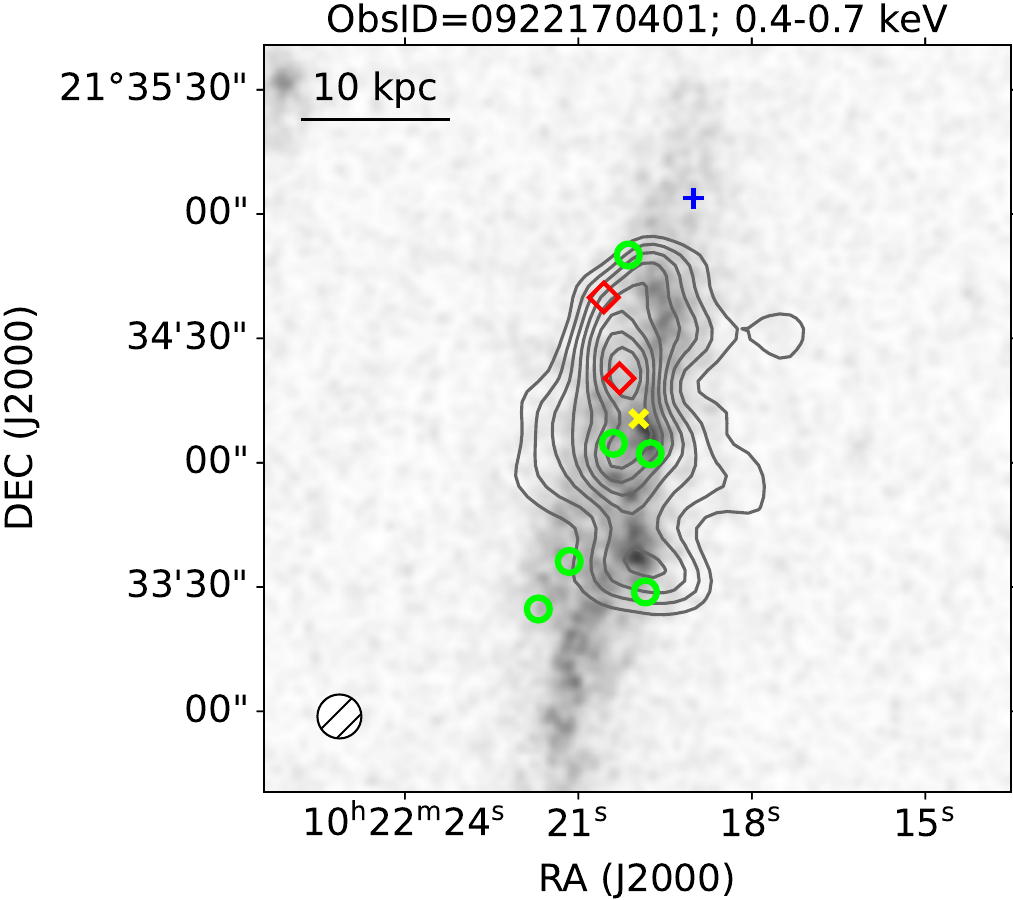}
    \includegraphics[trim=0 0 0 0, clip,width=0.195\linewidth]{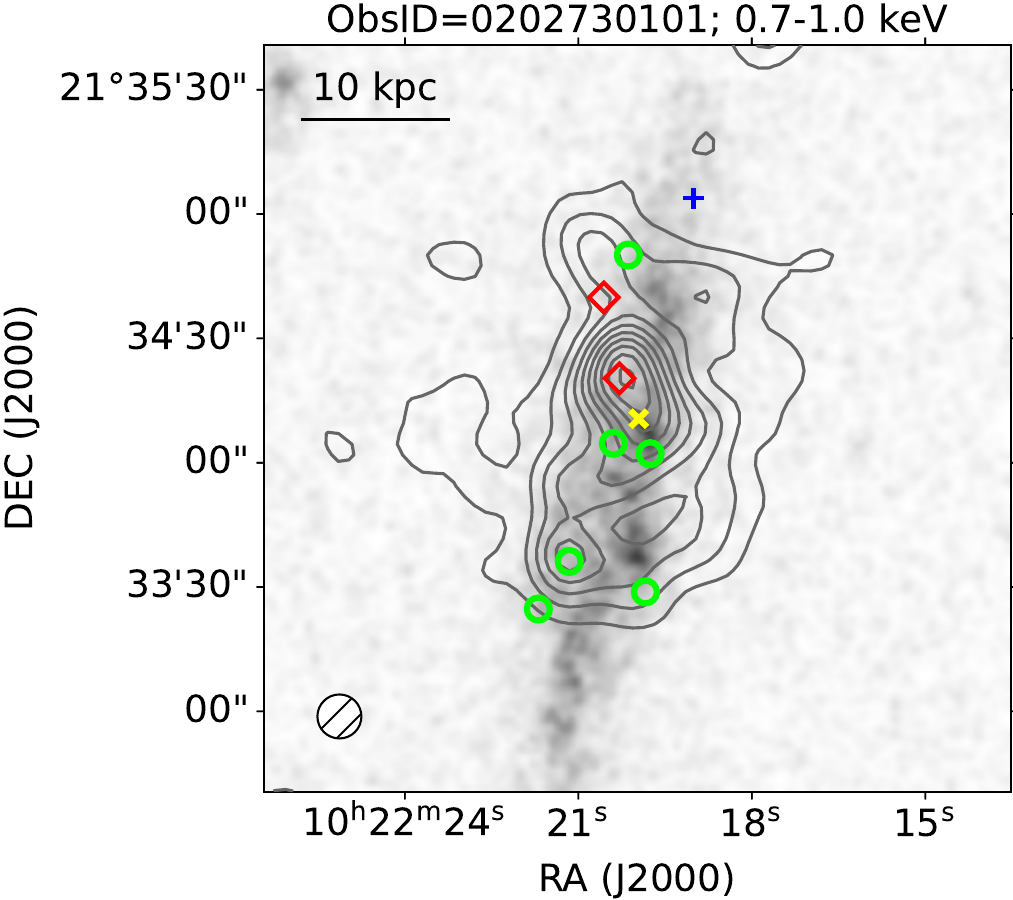}
    \includegraphics[trim=0 0 0 0, clip,width=0.195\linewidth]{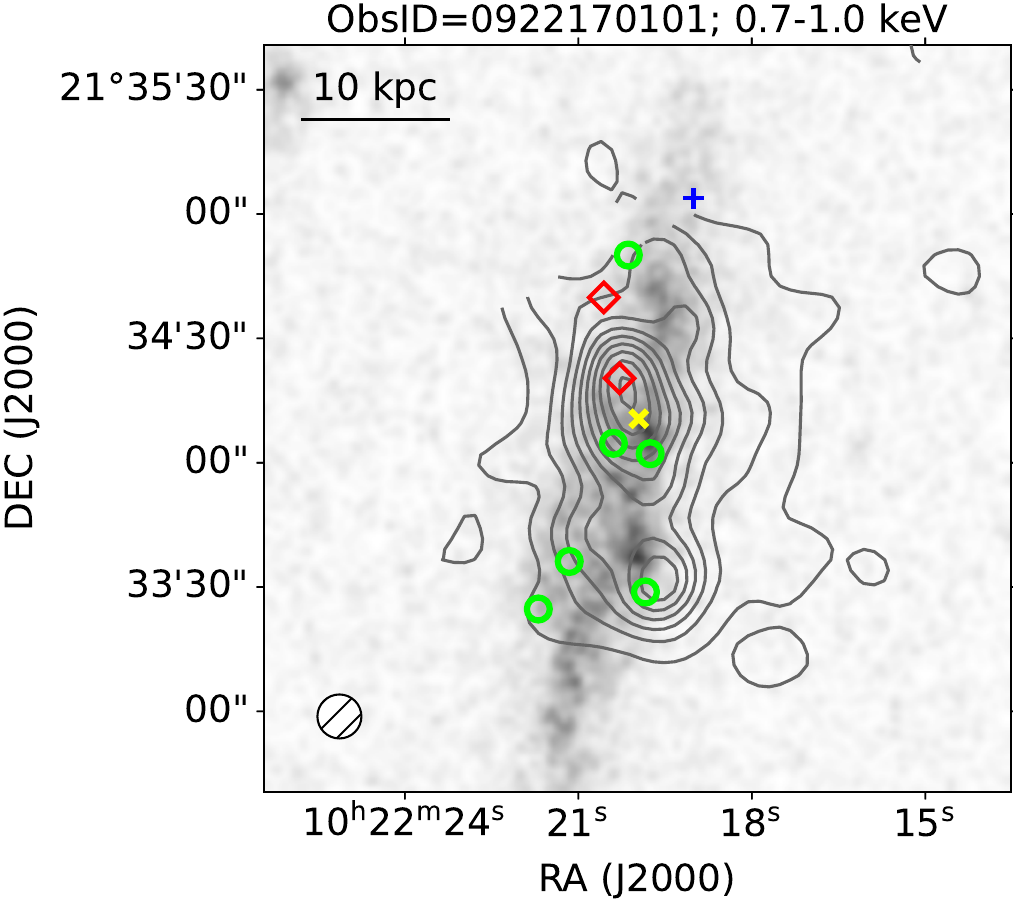}
    \includegraphics[trim=0 0 0 0, clip,width=0.195\linewidth]{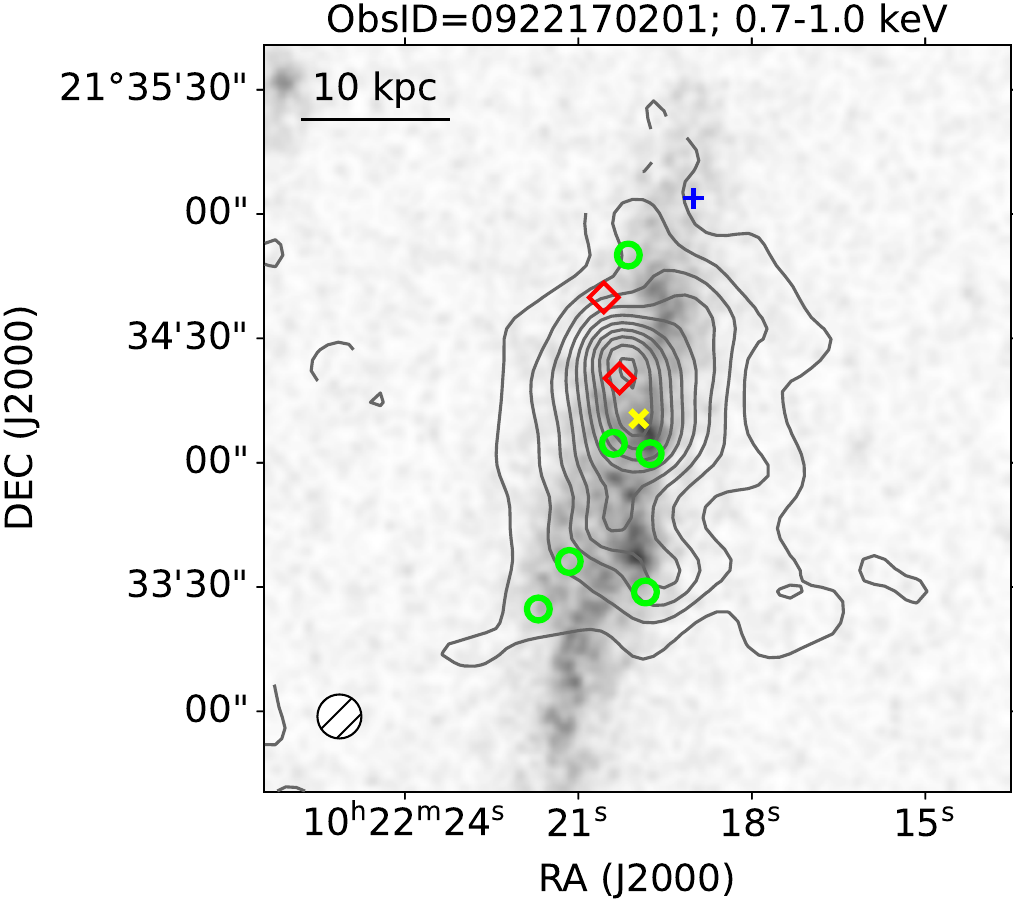}
    \includegraphics[trim=0 0 0 0, clip,width=0.195\linewidth]{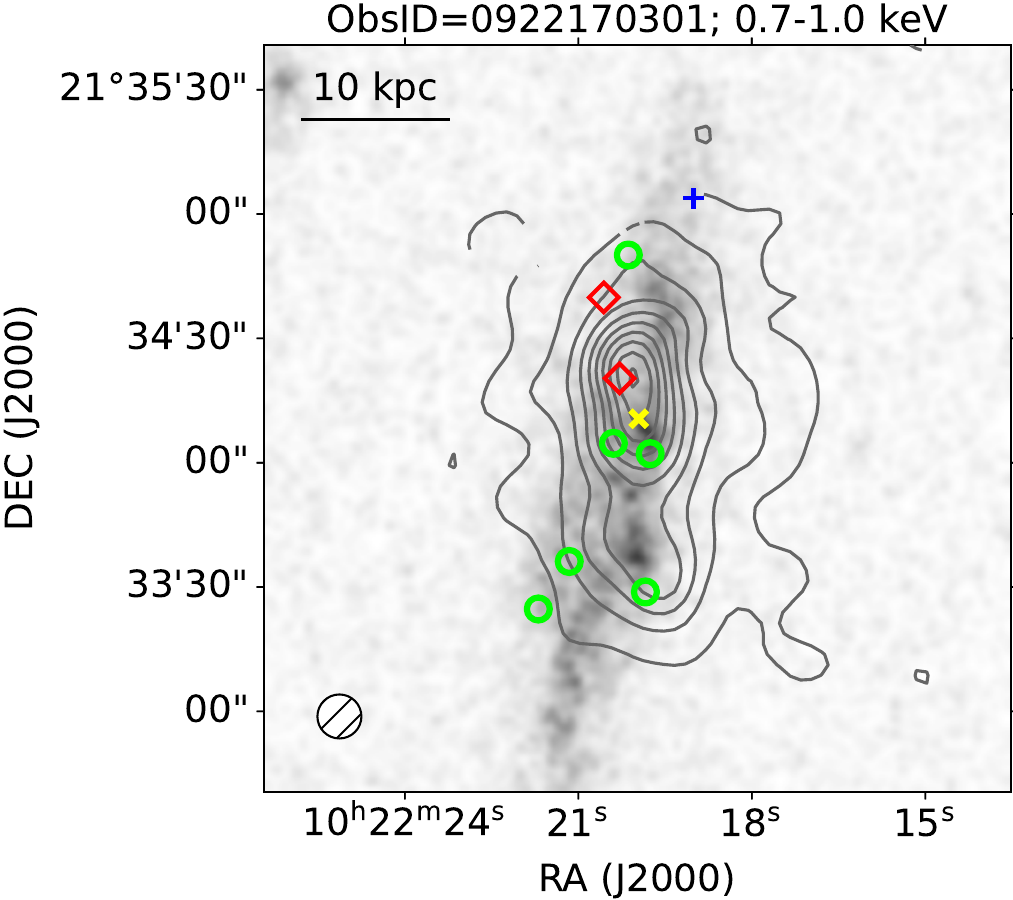}
    \includegraphics[trim=0 0 0 0, clip,width=0.195\linewidth]{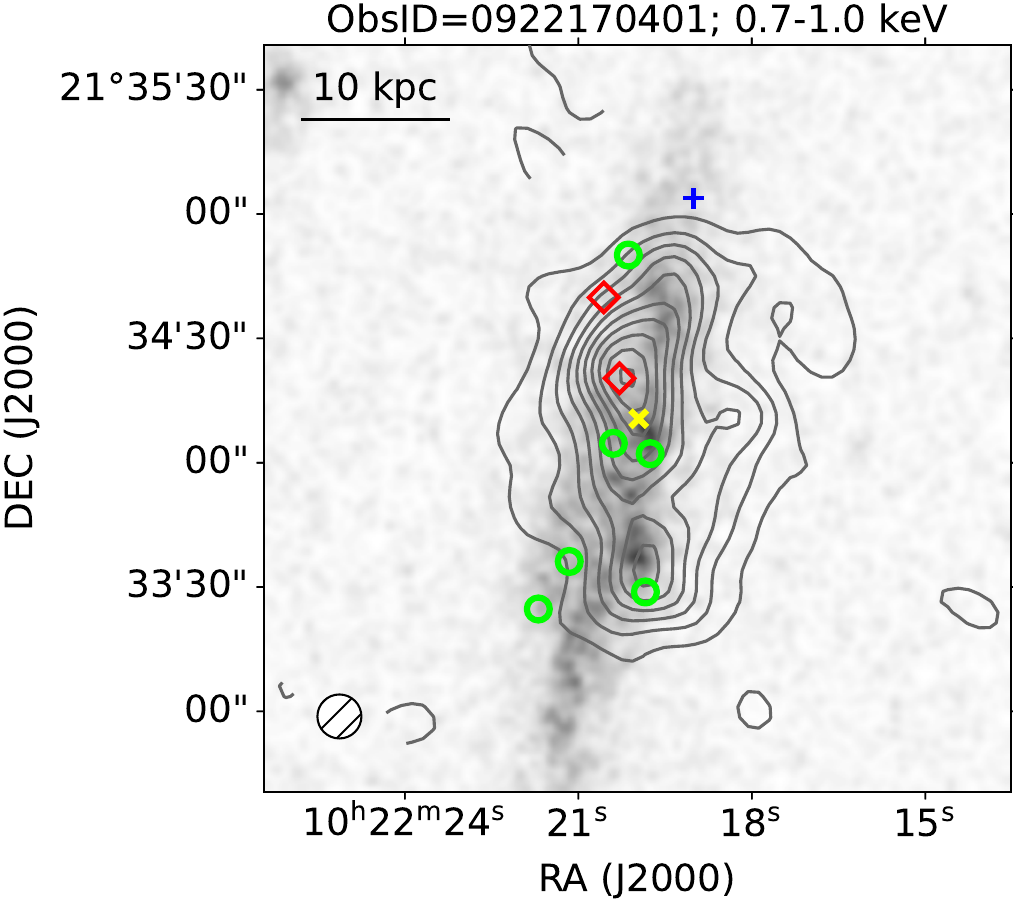}
    \includegraphics[trim=0 0 0 0, clip,width=0.195\linewidth]{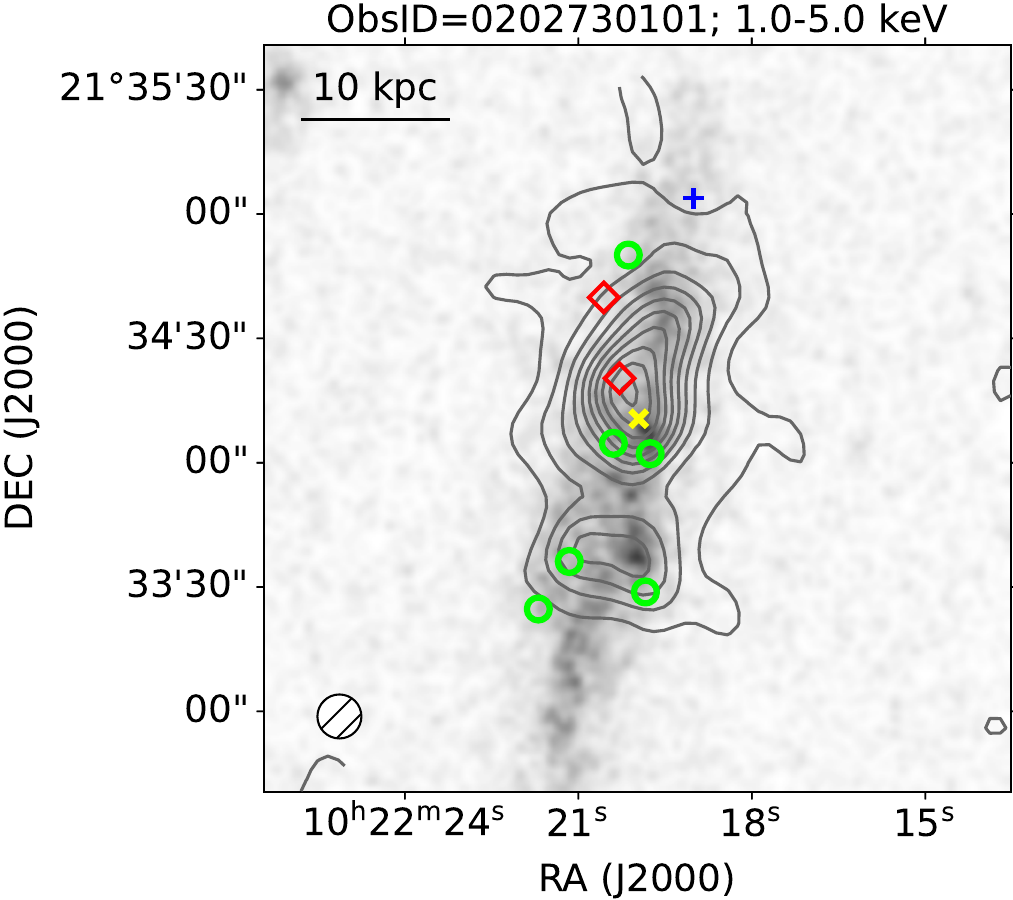}
    \includegraphics[trim=0 0 0 0, clip,width=0.195\linewidth]{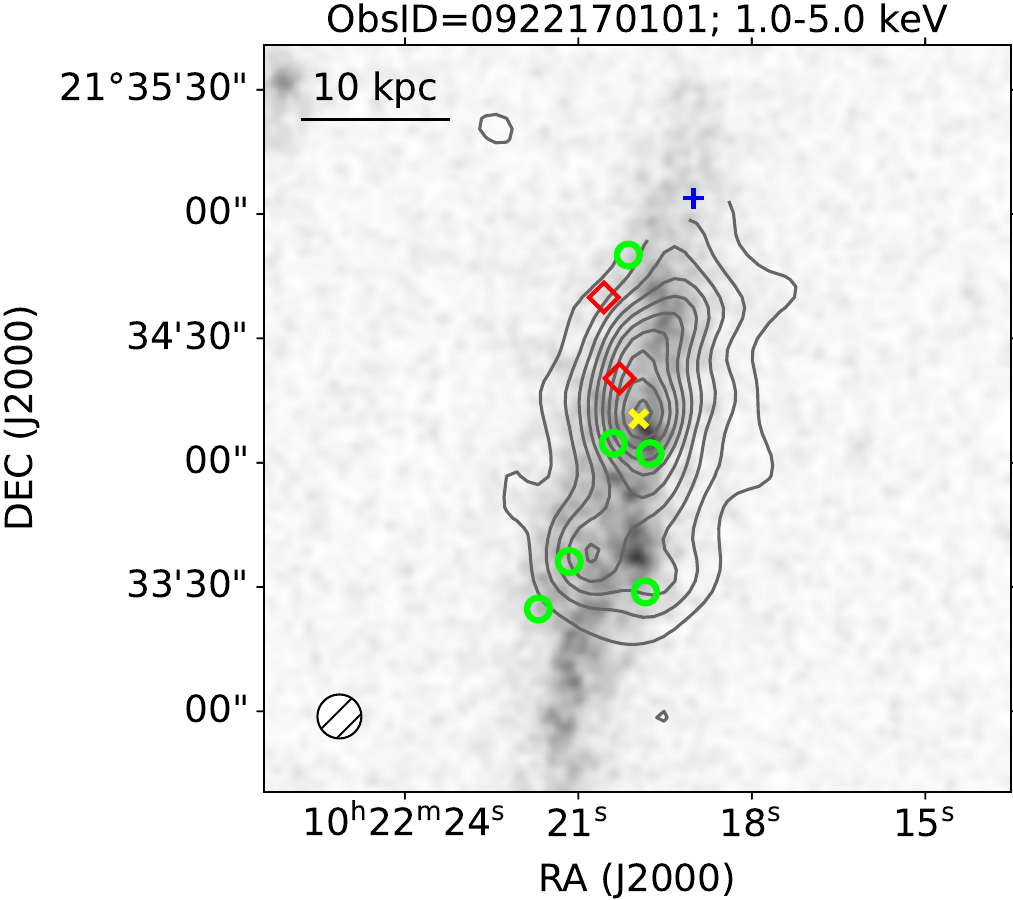}
    \includegraphics[trim=0 0 0 0, clip,width=0.195\linewidth]{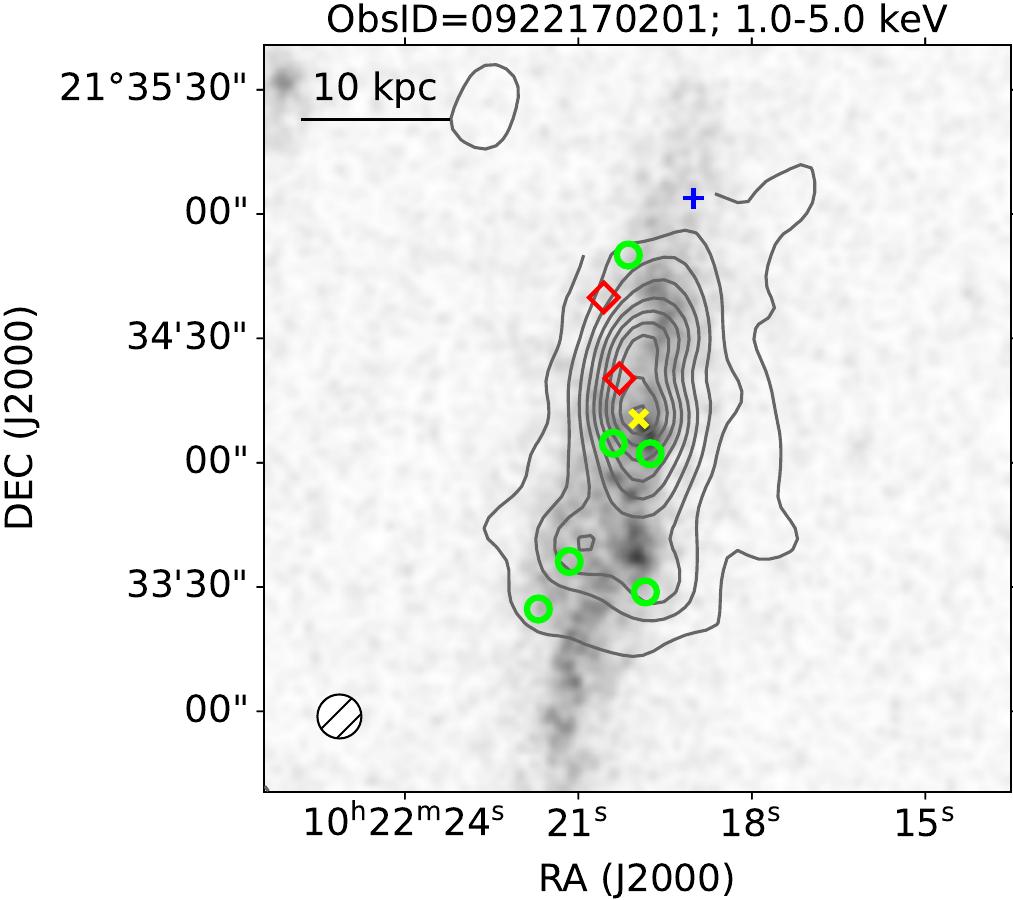}
    \includegraphics[trim=0 0 0 0, clip,width=0.195\linewidth]{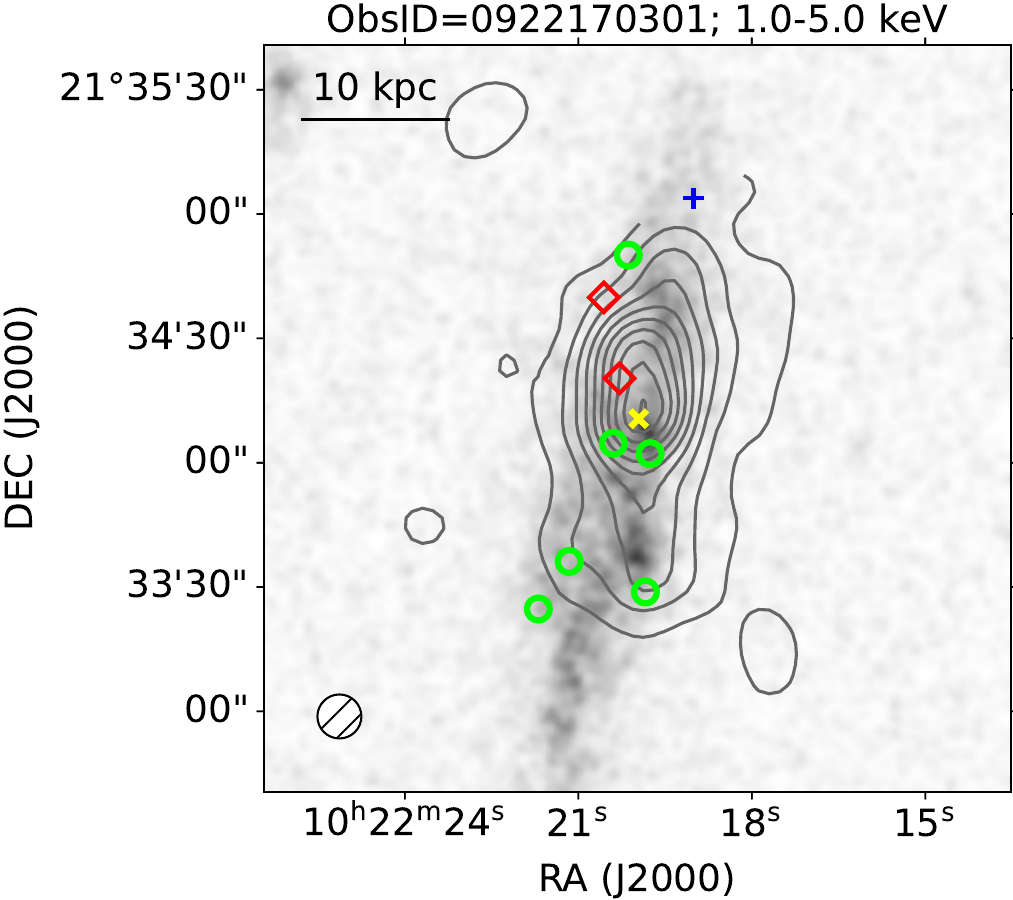}
    \includegraphics[trim=0 0 0 0, clip,width=0.195\linewidth]{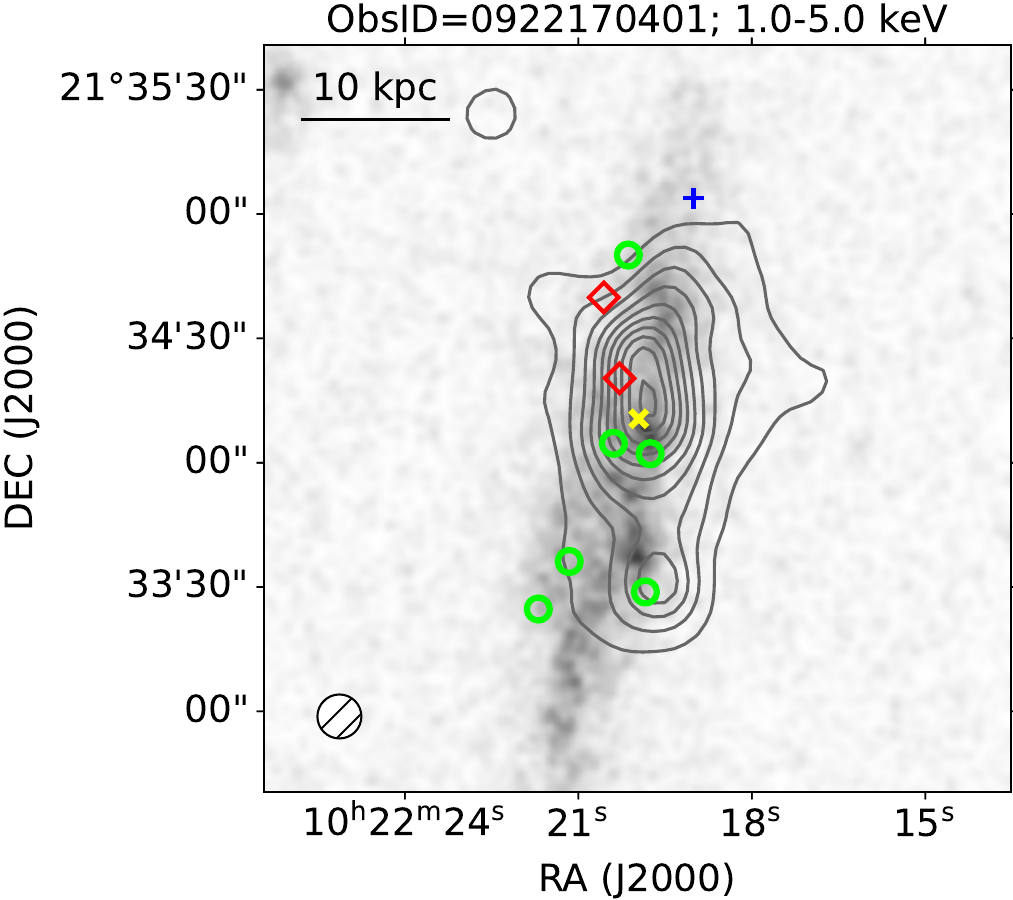}
    \includegraphics[trim=0 0 0 0, clip,width=0.195\linewidth]{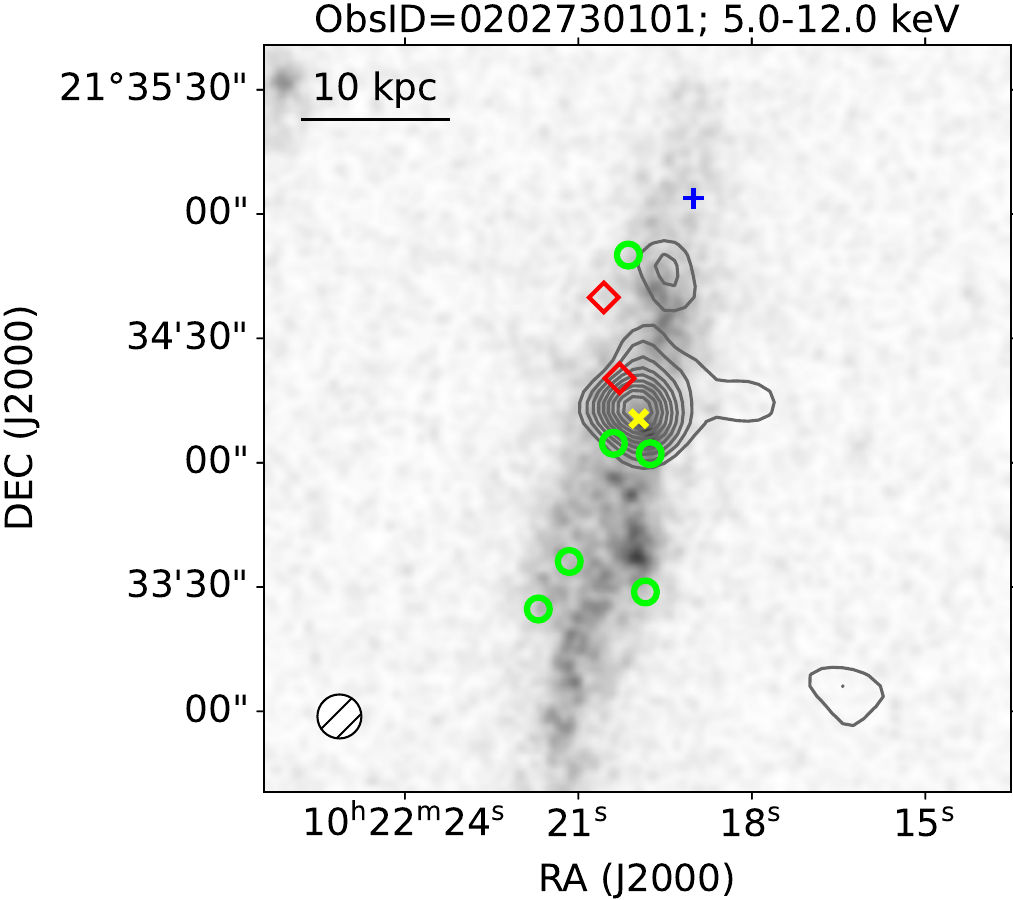}
    \includegraphics[trim=0 0 0 0, clip,width=0.195\linewidth]{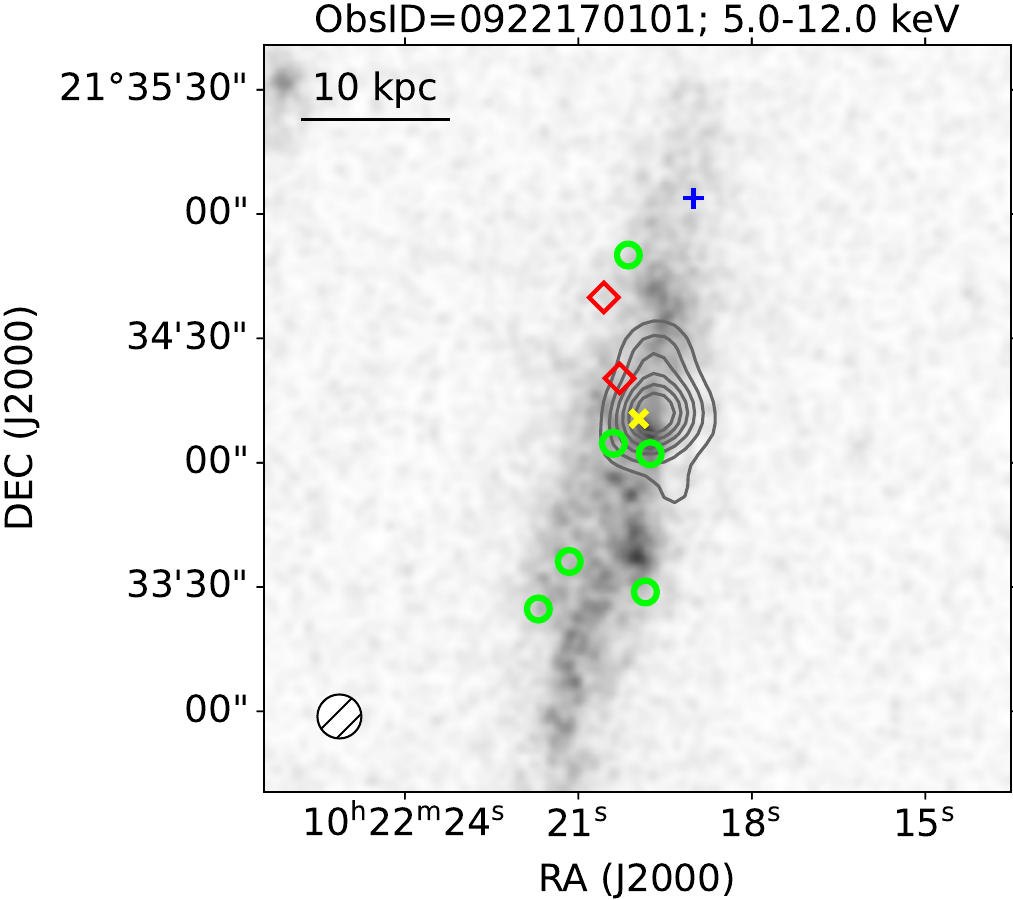}
    \includegraphics[trim=0 0 0 0, clip,width=0.195\linewidth]{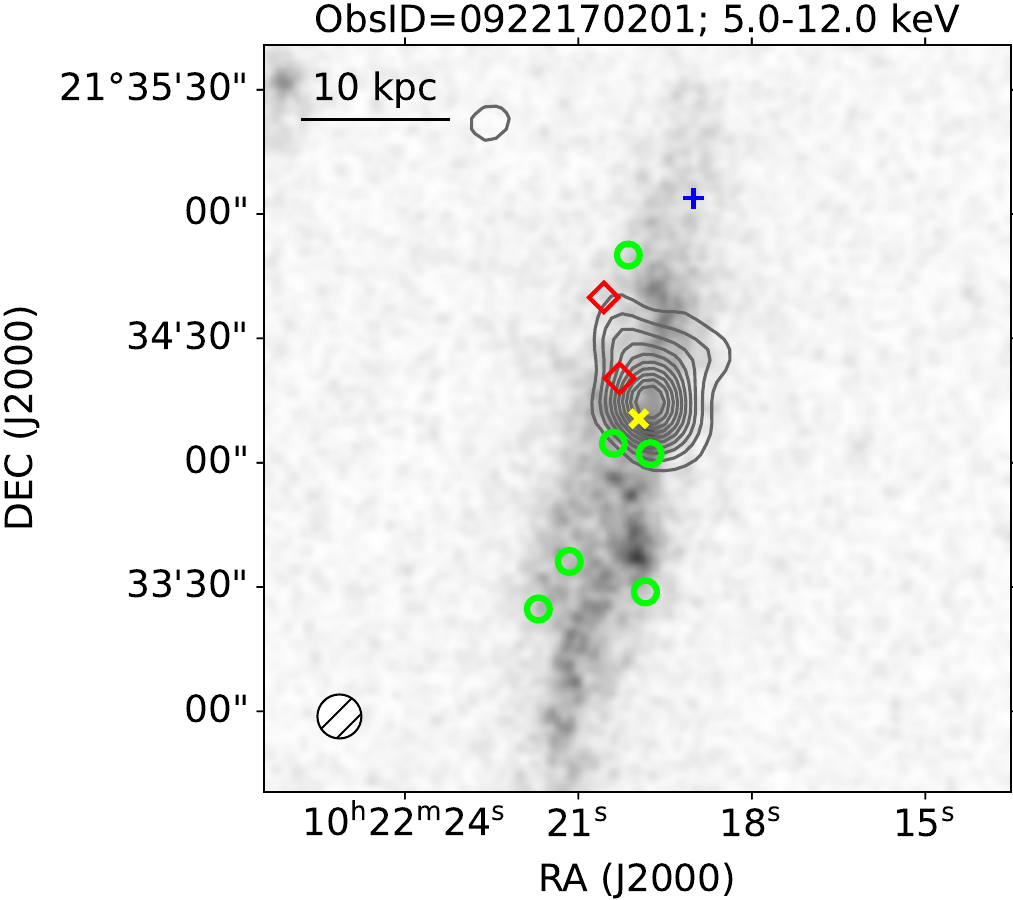}
    \includegraphics[trim=0 0 0 0, clip,width=0.195\linewidth]{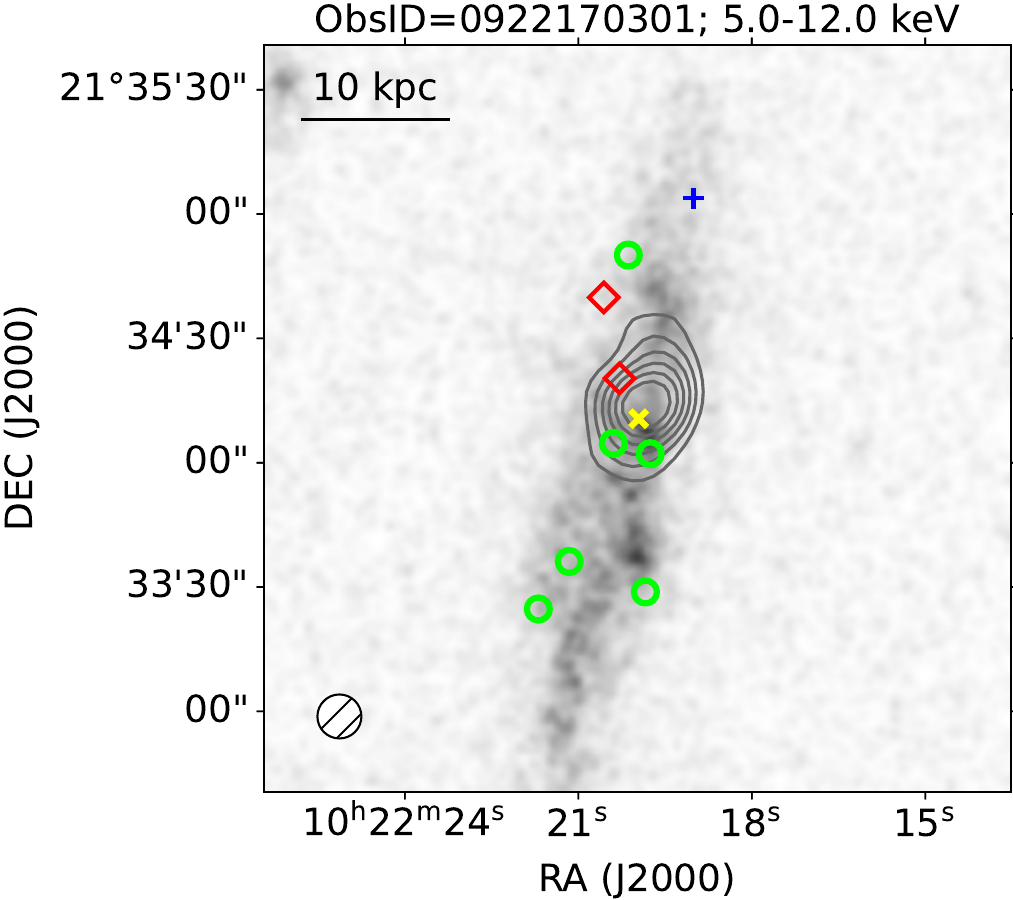}
    \includegraphics[trim=0 0 0 0, clip,width=0.195\linewidth]{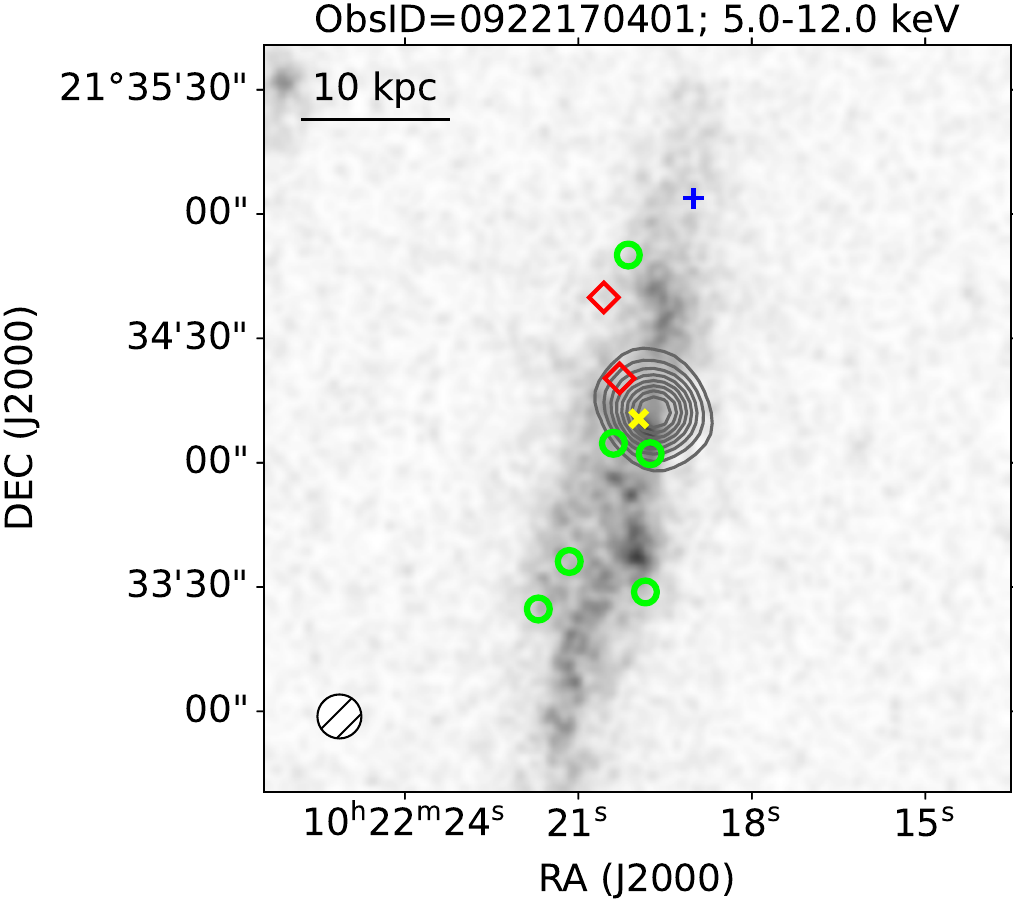}
    \includegraphics[trim=0 0 0 0, clip,width=0.195\linewidth]{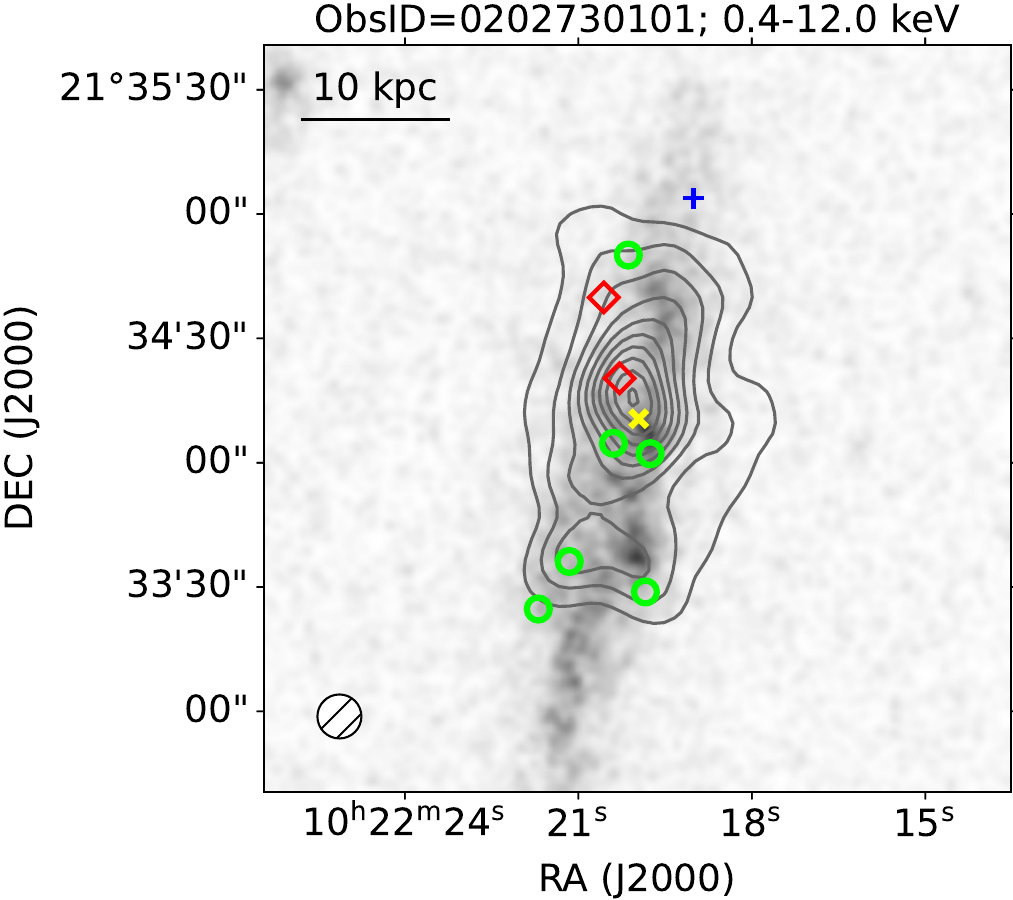}
    \includegraphics[trim=0 0 0 0, clip,width=0.195\linewidth]{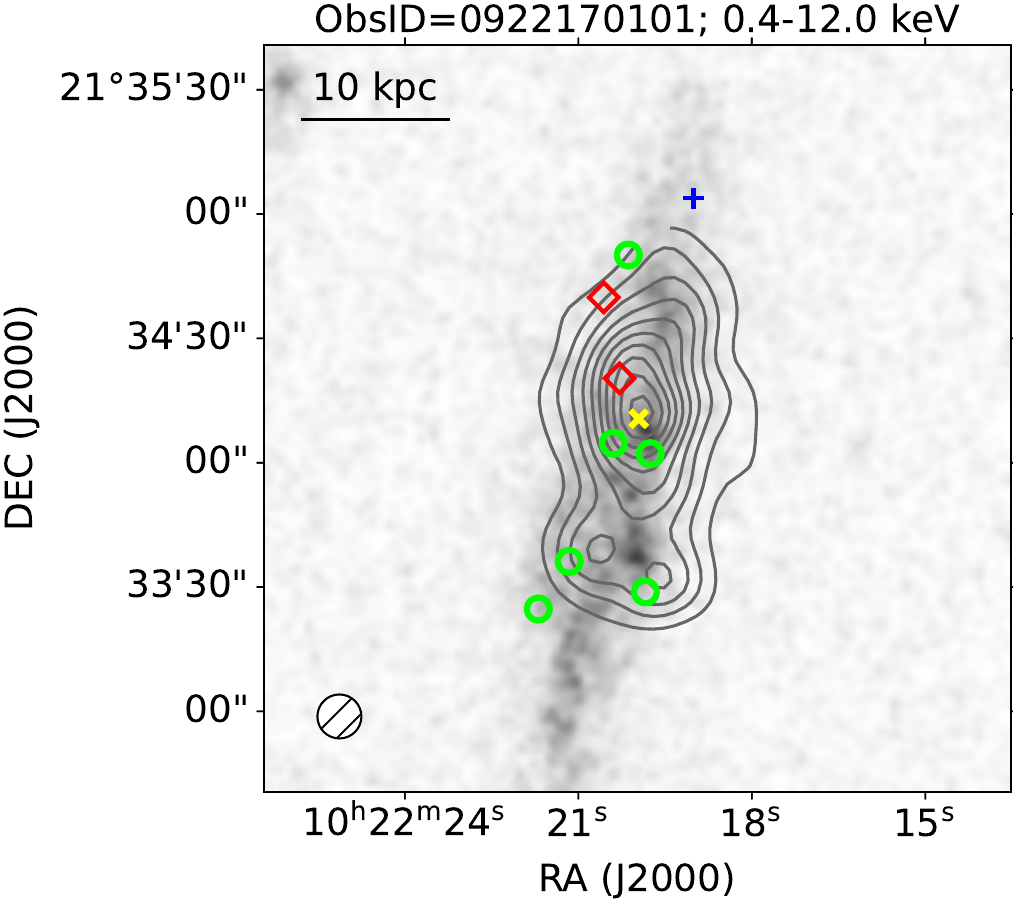}
    \includegraphics[trim=0 0 0 0, clip,width=0.195\linewidth]{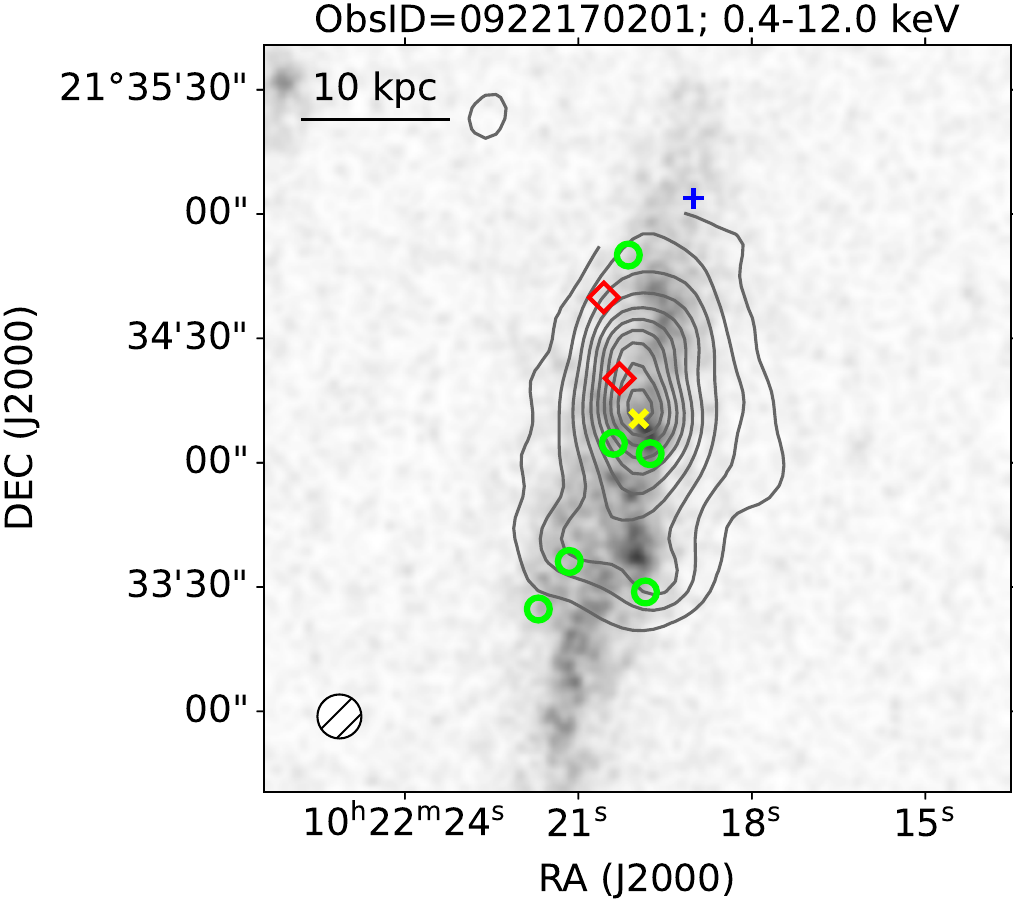}
    \includegraphics[trim=0 0 0 0, clip,width=0.195\linewidth]{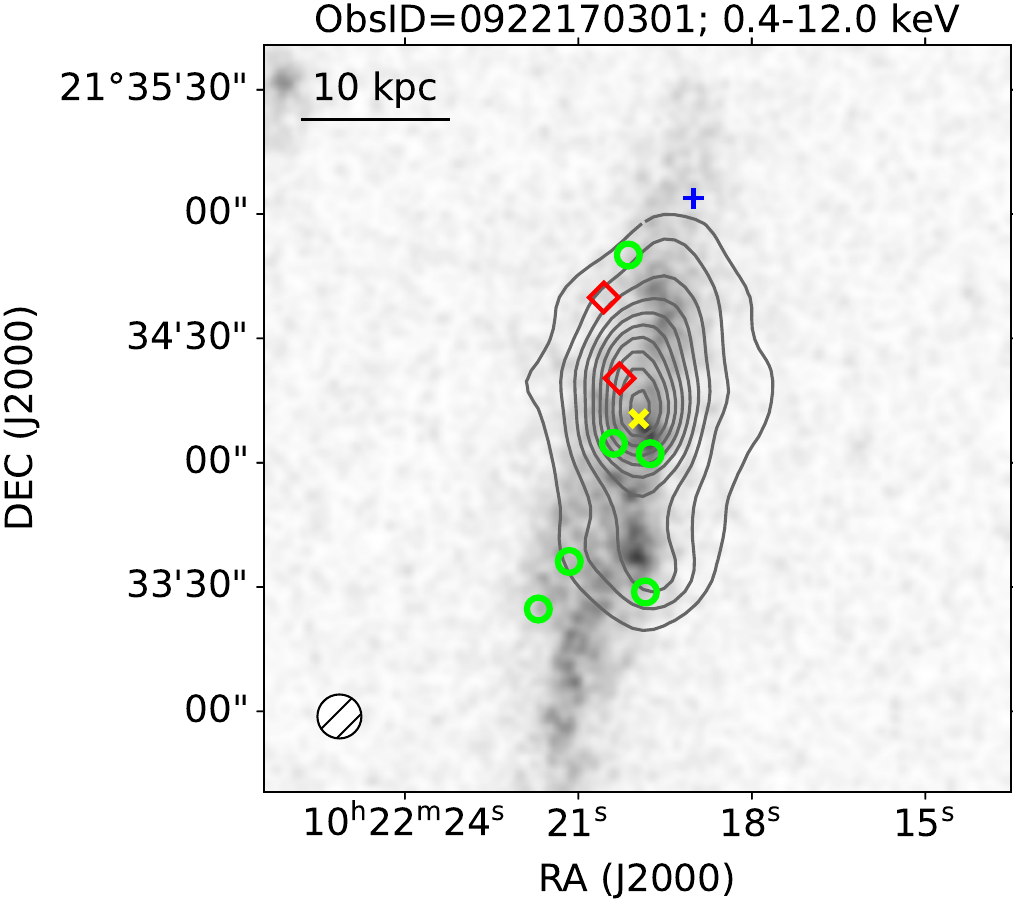}
    \includegraphics[trim=0 0 0 0, clip,width=0.195\linewidth]{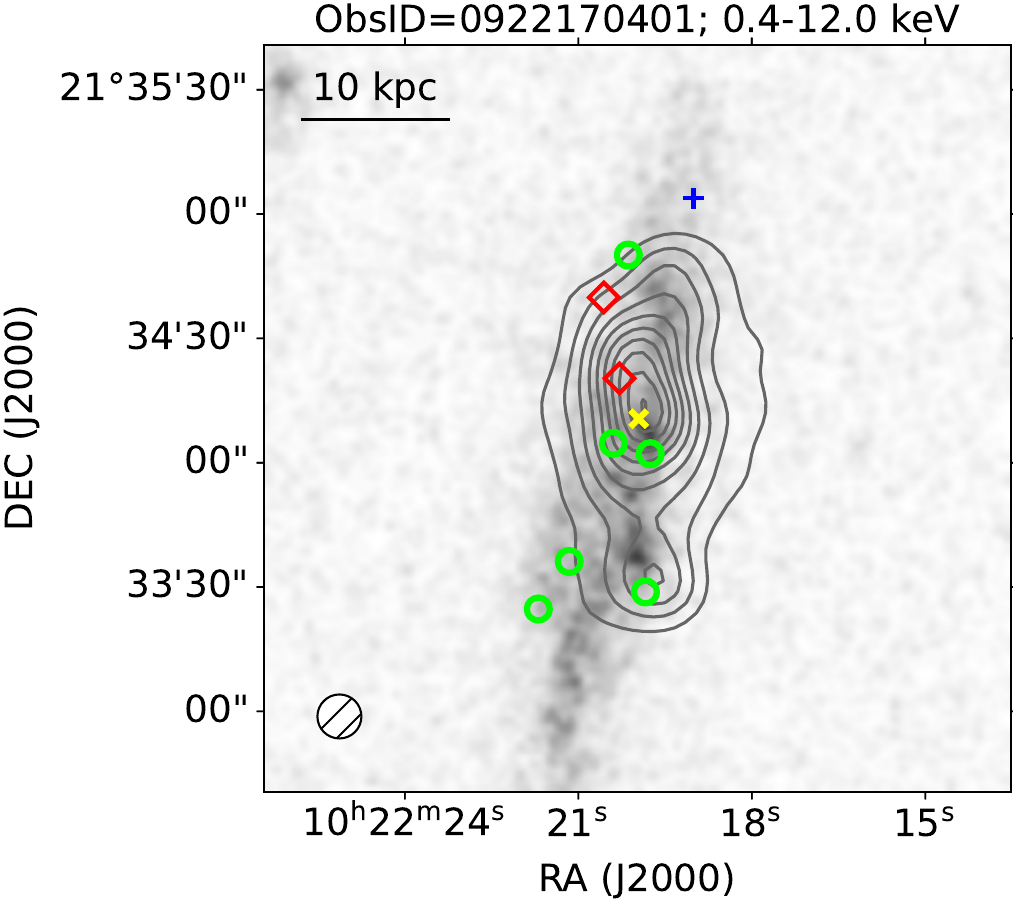}    
    \caption{\textbf{Top to bottom:} supersoft (0.4--0.7\,keV), soft (0.7--1.0\,keV), medium (1-5\,keV), hard (5-12\,keV) and broadband (0.4-12\,keV) contours from \xmmn/EPIC-pn data overplotted on GALEX-NUV images of \targetn. Contours are constructed after subtracting OOT and QPB and correcting for vignetting. Contours are drawn on a linear scale, the outermost contour is 3$\sigma$ above the background, and increases by 1$\sigma$ inside. \textbf{Left to right:} different obsIDs are shown chronologically, with the earliest to the left. Six point sources detected in the \chandra data and identified as ULX in \citetalias{Luangtip2015} are shown here with green circles. Blue `+' symbol denotes SN\,1961L. Yellow `x' marks the optical/IR center of the galaxy, i.e., the AGN. The red diamonds are drawn at the positions of two previously unpublished sources that are not detected in the \chandra data. The PSF of EPIC-pn is shown with a hatched circle at the bottom-left corner of each image. The scale of 10\,kpc is shown at the top left corner.}
    \label{fig:contour}
\end{figure*}

The contours from OOT- and particle background-subtracted images for all obsIDs and energy ranges are shown in Figure\,\ref{fig:contour}. Because the sky background is unlikely to vary across the shown region, the shape of the contours should represent the spatial distribution of the point sources and diffuse medium in the disk and extraplanar region of \targetn. 


\paragraph{Supersoft (0.4--0.7\,keV)}
It probes the highly ionized oxygen (\ovii and \oviiin) emitting regions and is expected to be dominated by diffuse warm-hot medium. However, some concentrated areas are brighter than the rest of the disk (Figure\,\ref{fig:contour}, {top row}) - it could be point sources with soft excess or denser regions of the warm-hot ISM. Some of these bright regions are not identified in all obsIDs, implying significant flux contributions from variable compact objects.

\paragraph{Soft (0.7--1.0\,keV)}
This probes the Fe L-shell emitting regions and is expected to be dominated by diffuse hot medium. Again, some concentrated regions are brighter than the rest of the disk (Figure\,\ref{fig:contour}, {2nd row from top}), but not necessarily at the same position as the supersoft sources. It implies different temperatures across the hot ISM, or similar temperatures but different levels of ISM obscuration in soft X-rays. Some of these bright regions contain compact objects indicated by their non-detection (hence, variability) in some obsIDs. The diffuse emission in the soft band is more extended above the disk than the supersoft emission. Because of relatively weaker obscuration above the disk, this difference indicates that the average temperature of the highly ionized gas in the extraplanar region could be higher than the ISM. We explore this possibility further in \S\ref{sec:spec_ism}.

\paragraph{Medium (1--5\,keV)}
This primarily probes the emission from resolved and unresolved compact objects, which would ideally be spread across the disk; it is consistent with the overall shape of the contour (Figure\,\ref{fig:contour}, {3rd row from top}). There are a few peaks in the contour that could be due to the brightest point sources. The similarity between the broadband (0.4--12\,keV) contours (Figure\,\ref{fig:contour}, {bottom row}) and the medium band indicates that the total count from the disk is dominated by the medium band. It is further verified in the next section on the point sources as well.     

\paragraph{Hard (5--12\,keV)}
Only one source at the center of the galaxy, likely the AGN, is bright in this energy (Figure\,\ref{fig:contour}, {second row from bottom}). This relatively higher hardness of the AGN than other point sources in the disk indicates that the AGN could be obscured. We explore this further in the following sections.    
\subsection{Temporal variation}\label{sec:time}
The point sources we consider are the six ULXs detected in the \chandra data \citepalias{Luangtip2015}, and the AGN. Although not a compact object, we consider SN\,1961L as a ``point" source because it is unresolved in the PSF of EPIC. We discuss two more intriguing regions brighter than the rest of the diffuse ISM in $<1$\,keV; we call them ``new" sources. We show the position of all these sources in Figure\,\ref{fig:contour} superposed on the contours. 

The contribution from OOT is subtracted depending on the mode of observation (full-frame=6.3\%, and extended full-frame=2.3\%). The counts from particle background are calculated from the QPB image at the position of each source. The counts from photon background in the background region (as defined before in \S\ref{sec:reduction_xmm}, step\#7) are calculated by subtracting the QPB image from the OOT-subtracted science image, and then adjusted for vignetting to estimate the photon background at the position of each source. The OOT-, photon background-, and particle background-subtracted counts are divided by vignetting-corrected exposure to calculate the count rate. We repeat this procedure separately for supersoft, soft, medium, hard, and broadband.

\begin{figure*}
\centering
\includegraphics[width=0.99\linewidth]{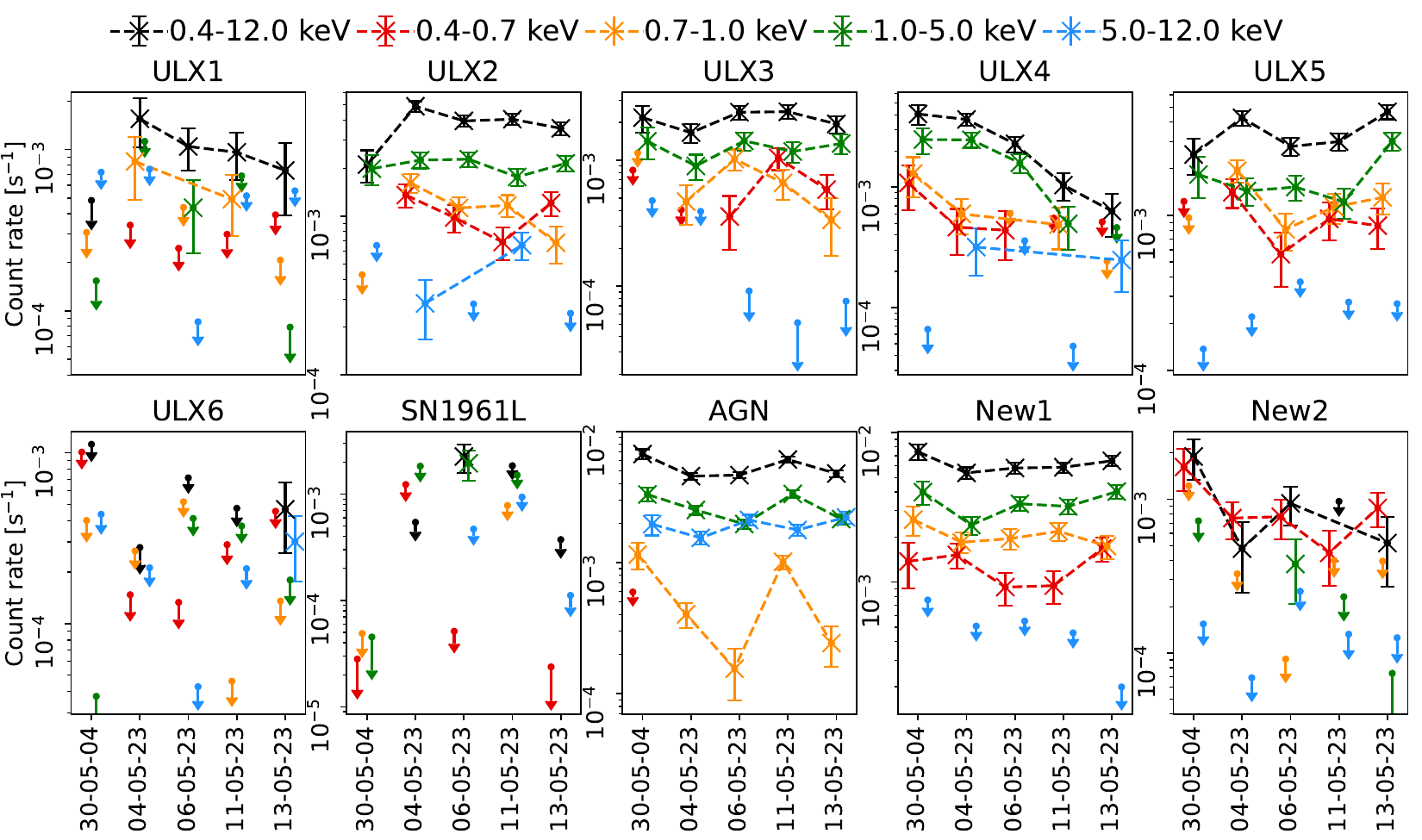}
\includegraphics[width=0.495\linewidth]{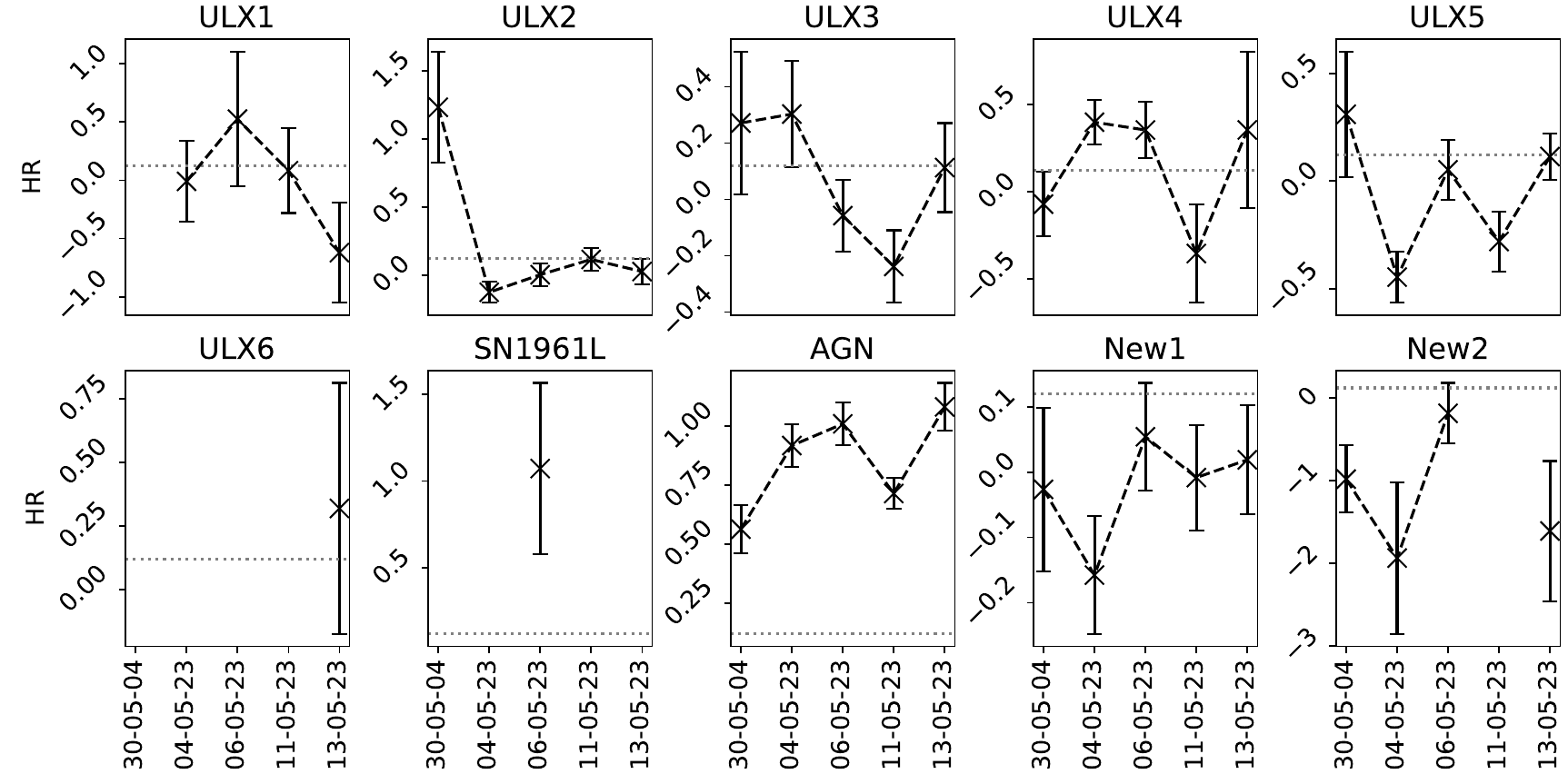}
\includegraphics[width=0.495\linewidth]{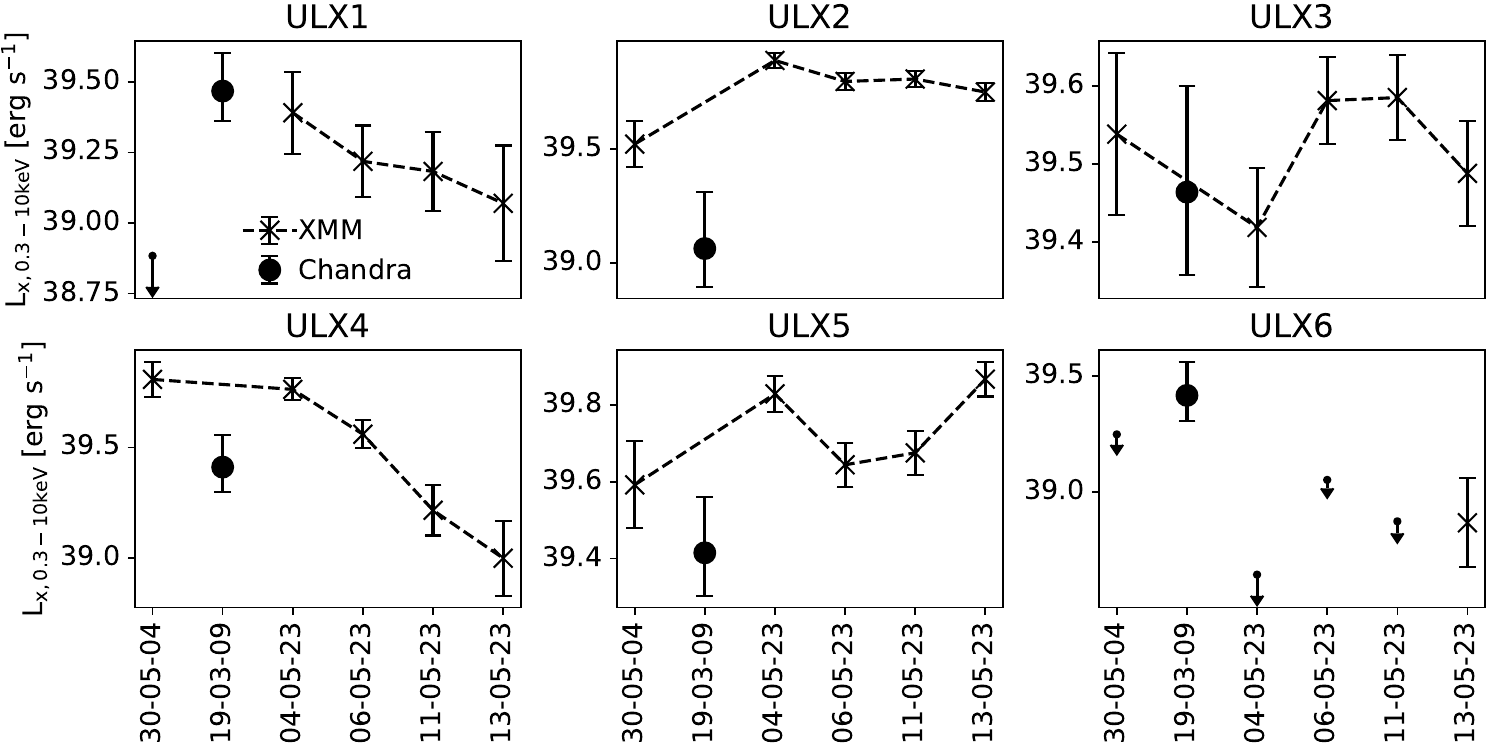}
    \caption{{Top:} Count rate and {bottom left (ten panels)} hardness ratios of the point sources from \xmm data, and {bottom right (six panels)} unabsorbed 0.3--10\,keV luminosity of the ULXs from \xmm and \chandra data. {The position of each object has been shown in Figure\,\ref{fig:optical} and \ref{fig:contour}.} The count rates for supersoft (0.4--0.7\,keV), soft (0.7--1.0\,keV), medium (1--5\,keV), hard (5--12\,keV), and broad (0.4--12\,keV) bands are displayed in red, orange, green, blue, and black, respectively. The hardness ratio for the SED that is assumed to estimate the luminosity is shown with gray horizontal dotted lines.}\label{fig:rate}
\end{figure*}

Because of the large PSF of EPIC-pn and the crowded environment of the disk, there could be residual contamination at the position of a point source by the surrounding point sources. Thus, for $n$ point sources, the measured count rate of the $i-$th object in the previous step, $c_{obs,i}$, is essentially $c_{obs,i} = \sum_{j=0}^{j=n} f_{i,j} c_{true,j}$, where $c_{true,j}$ is the \textit{true} count rate of the $j-$th object, and $f_{i,j}$ is the fractional contribution by the $j-$th source at the position of the $i-$th source. To estimate the contamination matrix $F$, we construct PSF maps in supersoft, soft, medium, hard, and broadband using \texttt{psfgen} for level=ELLBETA. This produces the most accurate map with a 2-D King function, including instrumental distortions and the spokes.   We obtain the \textit{true} count rates from the measured count rates using equation\,\ref{eq:trueL}, and plot in Figure\,\ref{fig:rate}, top).

\begin{equation}\label{eq:trueL}
\begin{split}
C_{true} = F^{-1}C_{obs} \\
{\rm where\;} C_{true} = 
\begin{bmatrix}
     c_{true,0}  \\
     \vdots& \\
     c_{true,n}
\end{bmatrix},\; C_{obs} = 
\begin{bmatrix}
     c_{obs,0}  \\
     \vdots& \\
     c_{obs,n}
\end{bmatrix}
\\
{\rm and\;}
F = 
    \begin{bmatrix}
f_{0,0} & f_{0,1} & ... & f_{0,n-1} & f_{0,n} \\
\vdots & &  \ddots &  & \vdots\\
f_{n,0} & f_{n,1} & ... & f_{n,n-1} & f_{0,n} \\
\end{bmatrix}
\end{split}
\end{equation}

Same as \citetalias{Luangtip2015}, we assume that the SED of ULXs follows an absorbed powerlaw of $\Gamma = 2$ with N(\hin)$\rm_{NGC\,3221} = 1.5\times 10^{21} cm^{-2}$. We adopt N(\hin)$\rm_{Milky\,Way} = 1.77\times 10^{20} cm^{-2}$ toward the sky direction of \target from \cite{Bekhti2016}. We calculate the unabsorbed 0.3--10\,keV luminosity from the observed absorbed 0.4--12\,keV count rate using PIMMS. The resulting luminosities are compared with those measured from \chandra data \citepalias{Luangtip2015} in Figure\,\ref{fig:rate}, bottom right).  

We calculate the hardness ratio as HR = (H-S)/(H+S), where H and S are count rates in 1-12\,keV and 0.4-1\,keV. The HR for the epochs with constrained count rates is shown in Figure\,\ref{fig:rate}, bottom left). The hardness ratio for the assumed SED for ULXs is shown with horizontal lines. From this, we verify whether the \textit{true} SED is variable, and whether it differs from the assumed SED. 

Below, we discuss every source separately. We numerically label the ULXs in descending order of the DEC, with the ULXs at the highest and lowest DEC being called ULX1 and ULX6, respectively.  We call it a detection if S/N$>$2, and otherwise provide a 2$\sigma$ upper limit.

\paragraph{ULX1} 
It is detected in all the new \xmm observations but not in the archival data. It continues to get fainter from the first new observation to the subsequent observations by a factor of $\approx 3$. It does not have any detectable emission in supersoft and hard bands. The hardness ratio during the first three new observations is consistent with the assumed SED. During the last observations, the source became softer, suggesting a variable soft excess. Its luminosity in all new \xmm data is smaller than that in the \chandra data, and the upper limit from the archival \xmm data is smaller than all detections. 

\paragraph{ULX2}
It is detected in all datasets. Its broadband count rate remains almost constant during the new \xmm observations and is larger by a factor of $\approx$2 than the archival data. The broadband count is dominated by the medium band. It has comparable emission in the supersoft and soft bands. In the first and the third new \xmm observations, it has detectable emission in the hard bands. The narrowband count rates vary by a factor of a few across observations, resulting in a variable hardness ratio. In the archival \xmm data, it was distinctly harder than the assumed SED of ULXs. In the new observations, it is closer to the assumed SED of ULXs, but becomes harder from the first to the fourth observation. It is more luminous in all \xmm data than in the \chandra data. 

\paragraph{ULX3}
It is detected in all datasets. Its broadband count rate slowly increases during the new \xmm observations that spanned a week, implying a longer variability timescale than ULX2. The broadband count is dominated by the medium band. It has comparable emission in the supersoft and soft bands, but no detectable emission in the hard band. Its narrowband count rates vary by a factor of a few, resulting in a hardness ratio varying on both softer and harder sides of the assumed SED of ULXs. Because the broadband count rate remains almost the same across observations, the varying hardness ratio implies a redistribution of its total energy among the soft and hard components. The luminosity of ULX3 during \xmm and \chandra observations is similar. 

\paragraph{ULX4}
It is detected in all datasets, but its broadband count rate monotonically decreases by almost an order of magnitude during the new \xmm observations. The supersoft (soft) band emission is detected in the archival \xmm and the first two (three) new \xmm observations. The medium band dominates its broadband count. In the first and the fourth new \xmm observation, it has detectable emission in the hard band. The count rate in different narrow bands varies significantly, resulting in a hardness ratio varying between a large range of $\approx -0.5$ to 0.6. Its luminosity in the archival and the first new \xmm data is larger, and in the last two \xmm data is smaller than that in the \chandra data. 

\paragraph{ULX5}
It is detected in all datasets. Its broadband count is dominated by the medium band in three observations and is equally contributed by softer bands in the other two observations (0922170101 and 0922170301). It has no detected emission in the hard band; the supersoft and soft band emission are of comparable strength and are detected in all the new \xmm observations (but not in the archival data). Its luminosity in all the \xmm data is larger than the \chandra data. Its hardness ratio oscillates between $\approx -0.5$ to $\approx 0.5$. The hardness ratio during 0922170101 and 0922170301 is smaller, while the other three \xmm observations are consistent with the assumed SED, suggesting a variable soft excess. Because the broadband count rate remains almost the same across observations, it implies a redistribution of its total energy among the soft and hard components.

\paragraph{ULX6}
It is detected in only one new \xmm observation. The hardness ratio is consistent with the assumed SED, albeit with a large uncertainty. The detected luminosity or the upper limits in the new  \xmm observations are $\gtrsim 0.7$\,dex lower than that in the \chandra data, showing high variability. 


\paragraph{SN\,1961L}
It is detected in one \xmm observation, with almost all of the emission from the medium band. For the same assumed SED of ULXs, the 0.3--10\,keV unabsorbed luminosity would be 3.6$\pm$1.1$\times 10^{39}$ erg s$^{-1}$. In the archival \xmm data it has a 2$\sigma$ upper limit of 2.7$\times 10^{38}$ erg s$^{-1}$. Because of its position close to the chip gap in the other three new observations, the upper limits from them are not sufficiently constraining. However, the hardness ratio is larger than that for the assumed SED by $>2\sigma$, implying the SED is distinctly different from an absorbed powerlaw. 

\paragraph{AGN}
It is detected in all datasets. The variable count rate in different narrow bands results in a hardness ratio varying between $\approx 0.5-1.1$. Given the HR, its SED should be different from that assumed for ULXs. We further discuss the SED of the AGN in \S\ref{sec:spec_agn}.

\paragraph{New source-1}
This source is present in the archival \xmm data (although not reported before) as well as all the new datasets, but not in the \chandra data. Had it been a purely diffuse structure, it would not be variable and would be detected in the \chandra data, given its current count rate. The non-detection in the \chandra data indicates that it has to be purely/partially a compact object. It could be blended with an unresolved star cluster(s) and/or a superbubble.

It is the brightest object in the disk in $<$1\,keV. The exact location of the source cannot be pinpointed in the \xmm data due to poor angular resolution of EPIC and the crowded environment of the disk of \targetn. We use the center of the highest contour in soft X-ray as the tentative location of this ``new" source.

Most of its emission comes from the medium band. Softer bands ($<$1\,keV) are the second largest contributor. In the archival and two new observations (0922170201 and 0922170301), supersoft emission is weaker than the soft emission; in the other two observations, emission in supersoft and soft bands has similar strength. It has no detected hard band emission. 

The varying count rate in different narrow bands results in a hardness ratio varying between $\approx -0.2$ to 0.2. But because the broadband count rate remains almost the same across observations, it implies a redistribution of its total energy among different components. We further discuss the SED of the ``new" source in \S\ref{sec:spec_new}.

\paragraph{New source-2}
It is the second brightest source (after New source-1) in the supersoft band. We identify its tentative location in the same way as the new source-1. It is detected in the archival (although not reported before) and three new \xmm observations, but not in the \chandra data and the one new \xmm observation. The non-detection in \chandra could be because of the source variability or the low point source flux sensitivity of the \chandra data in the supersoft band. From its variability in the count rate across \xmm observations, we infer that it has to be purely/partially a compact object. It does not have any detectable emission above 1\,keV. Its total emission predominantly comes from the supersoft band, resulting in a hardness ratio around $\approx-2 \;\rm to\; 0$.

\subsubsection{Implications}
The 0.3--10\,keV aggregated luminosity of all the detected point sources (except the AGN) is $3.2\pm0.4\times 10^{40}\rm erg\,s^{-1}$, with $\sim1/3$rd of it coming from the ``new" sources. This is consistent with the observed $\rm L_X-M_\star-SFR$ relation of nearby galaxies within scatter \citep[e.g.,][]{Lehmer2010,Mineo2012a}. 

The total luminosity of the ULXs averaged over all the \xmm data, $2.0\pm0.3\times 10^{40}\rm erg\,s^{-1}$, is consistent with that in the \chandra data of $1.5\pm0.2\times 10^{40}\rm erg\,s^{-1}$ \citepalias{Luangtip2015} within 1$\sigma$. However, including the ``new" sources, the \xmm luminosity is larger by a factor of $2.1\pm0.4$. 
Given the diverse variability timescale observed among these ULXs (Figure\,\ref{fig:rate}), the difference in \chandra and \xmm luminosity indicates that the brighter part of the X-ray luminosity function (XLF) in an individual galaxy can be highly variable. Thus, if multiple galaxies of similar properties (e.g., stellar mass, SFR, metallicity, age, etc.) are observed with the same flux sensitivity but at different states of their short-term X-ray binary (XRB) activity, their measured XLF could appear distinct from each other despite the expected XLF to be similar. Therefore, the observed XLF would be a stochastic sample of an underlying population with similar time-averaged properties. 
Secondly, in observations, the narrowband counts are often not sufficiently large to constrain the SED. Therefore, the XLF is calculated assuming the same SED for all XRBs. However, the hardness ratio of ULXs of \target indicates that the \textit{true} SED often differs from that assumption and also varies over time (Figure\,\ref{fig:rate}). Thus, we conclude that the diversity in the shape and variability in the normalization of XRB spectra might be the reason behind the discrepancy often found between the predictions from detailed binary population synthesis models and observations \citep{Lehmer2019,Misra2023}.  

\target is edge-on, so the contamination by cosmic X-ray background on the XLF \citep{Kim2007} is likely negligible due to large absorption in the stellar disk. This might also be the reason for the non-detection of normal (not ultra-luminous) XRBs.    

We have explored day-scale and decade-scale variability. Because each of the new \xmm observations is almost a day long, it is possible to obtain hour-scale light curves from them for a finer investigation and a better understanding of the physical characteristics of the point sources. This is beyond the current scope of the paper.

\subsection{Spectroscopy}\label{sec:spectroscopy}
The AGN, one of the new sources (New-1), and the diffuse medium have sufficient counts in the new \xmm datasets for spectroscopy.  We analyze their spectra in \texttt{xspec} under a Bayesian environment using the cstat fit statistic. The solar chemical composition is set to follow the prescription of \cite{Asplund2009}.

\subsubsection{AGN}\label{sec:spec_agn}
\begin{figure*}
\centering
\includegraphics[width=0.99\linewidth]{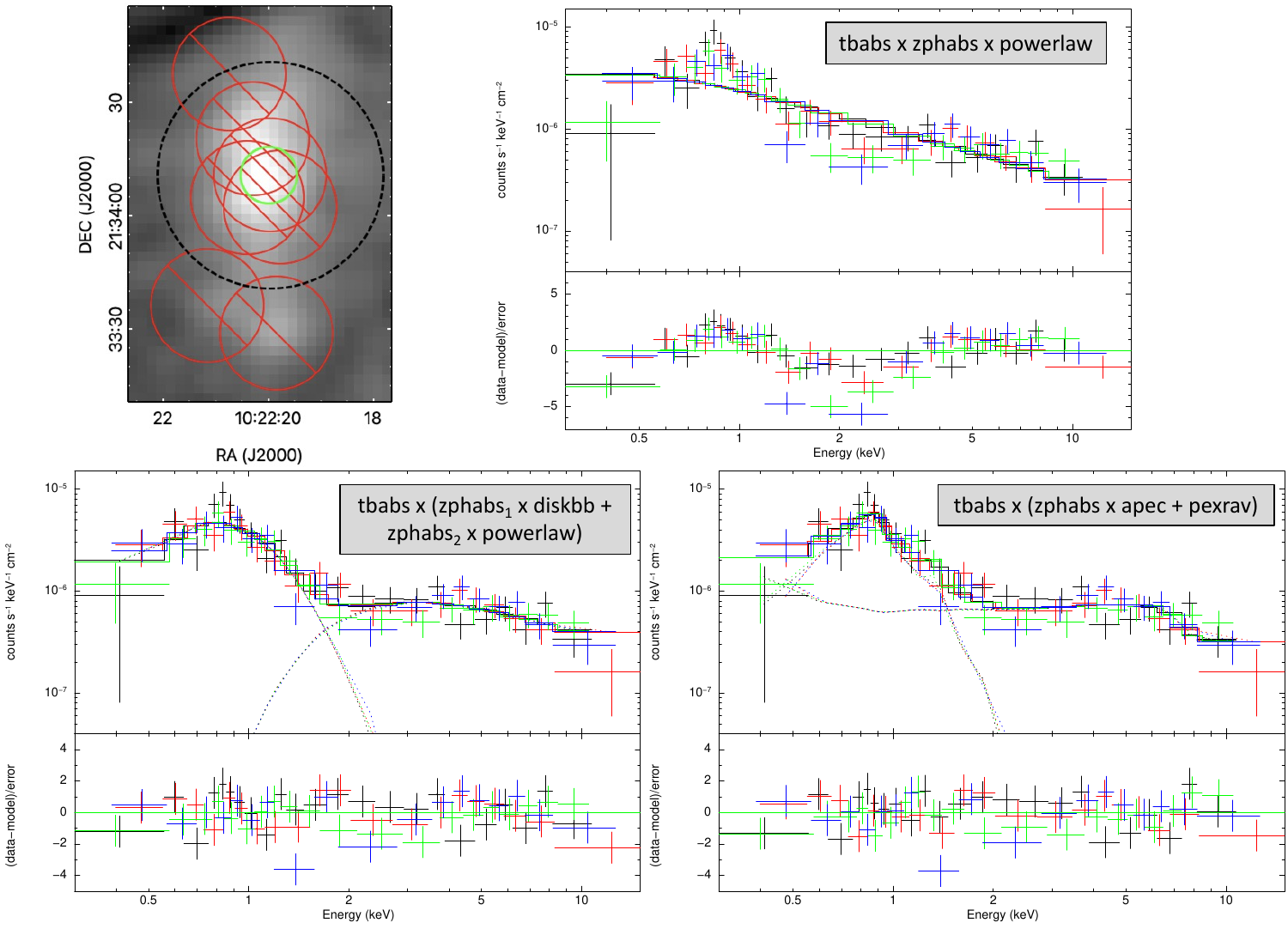}
    \caption{Top left: the extraction region for the AGN (``on-source" region) is shown in a green circle, and the background region is shown in a black dashed circle with the regions of red circles masked - plotted over the 0.4--12 keV science image of 0922170301. 0.4-12\,keV spectrum of the AGN is fitted with an absorbed powerlaw (and discarded; top right), and a combination of an absorbed disk blackbody and absorbed powerlaw (bottom left). A physically better (and statistically indistinguishable) alternative of an absorbed circumnuclear emission (\texttt{apec}) and the reflection of a Compton-thick AGN (\texttt{pexrav}) is shown in the bottom right. ObsIDs 0922170101, 0922170201, 0922170301, and 0922170401 are shown in red, green, black, and blue.}\label{fig:agn}
\end{figure*}
We fit the 0.4--12\,keV AGN spectra of four new obsIDs simultaneously with a powerlaw absorbed by the disk of Milky\,Way and the disk of \target and the local absorption around the AGN (\texttt{tbabs$\times$zphabs$\times$powerlaw}). The N(\hin) of \texttt{tbabs} is kept fixed, and the N(\hin) of \texttt{zphabs} is allowed to vary. The model does not fit the data well (cstat/dof = 203.39/80) with excess in the data at 0.7--1.1\,keV and deficit at $<$0.6\,keV 1.5--2.5\,keV (Figure\,\ref{fig:agn}, top right).

{The soft excess, often observed in AGN spectra, is described by a variety of models \citep[e.g., see][]{Mathur2017b}. As is often done in the literature, we parametrize the soft excess with an accretion disk with blackbody emission, \diskbbn.} We allow the intrinsic absorption of the powerlaw and blackbody to differ: \texttt{tbabs(zphabs$_1$$\times$diskbb + zphabs$_2$$\times$powerlaw}). It significantly improves the fit (cstat/dof = 80.16/77), requiring the additional component at 100\% confidence (ftest null hypothesis probability 1.5e-15; Figure\,\ref{fig:agn}, bottom left). The best-fit parameter values are provided in Table\,\ref{tab:bestfit}. The unabsorbed 0.2--10\,keV luminosity of the AGN in this model is $(4.2\pm 1.5)\times 10^{40} \rm erg s^{-1}$, with comparable contributions from the powerlaw and disk blackbody. To test if the data require different absorbing columns for the disk blackbody and powerlaw, we refit the spectra by tying the two \texttt{zphabs} components. The fit becomes worse (cstat/dof = 83.05/78), implying that the absorbing columns for the disk blackbody and powerlaw are different at 90.0\% confidence. 

Using the $M_\star-\sigma$ relation from \cite{Reines2015}, we derive the expected black hole mass to be $\rm \approx 10^{7.45}M_\odot$. For the best-fit values of the \diskbb parameters, the accretion efficiency, $\rm \dot{M}_{BH}/M_{BH}$, is $0.10_{-0.06}^{+0.14} (cos\,i_{acc})^{-1.5} \rm (M_{BH}/10^{7.45}M_\odot)^{-1}$, with $i_{acc}=0^\circ$ implying a face-on accretion disk. Because the stellar disk of \target is at $i =79^\circ$, the \textit{true} accretion efficiency could be higher, although misalignment between the accretion disk and stellar disk is not impossible. 

{The powerlaw photon index of AGNs ranges from $\Gamma=1.7$ to $\Gamma=2.4$ with a mean $\Gamma=1.95$ \citep{Liu2016}.} However, in the above analysis, the photon index of the powerlaw, $\Gamma = 0.9 \pm 0.2$, is unusually small and therefore is unlikely to be the \textit{true} photon index of the AGN. The AGN continuum can be flat if the direct continuum is completely obscured and only the reflected continuum is observed. Therefore, as a plausible alternative to the absorbed powerlaw model discussed above, we fit the hard component of the spectra with the reflected portion of a Compton-thick AGN featuring an exponentially cut-off powerlaw spectrum. In this scenario, the soft excess cannot come from the disk blackbody as the direct emission from the AGN is completely obscured. Faint, diffuse circumnuclear emission is often observed when the strong direct AGN continuum is heavily obscured \citep[e.g., type-2 AGNs like NGC\,1068;][]{Young2001}. Therefore, we replace the \diskbb model with a collisional plasma in thermal and ionization equilibrium to explain the diffuse circumnuclear emission. Thus, the alternate model we explore is \texttt{tbabs(zphabs$\times$apec + pexrav}). We assume that the accretion disk is aligned with the stellar disk{, i.e., \texttt{cosIncl}=cos(79$^\circ$),} and has a solar metallicity{, i.e., \texttt{abund}=1 and \texttt{Fe\_abund}=1}. Statistically, this Compton-thick AGN model fits the data as good as the previous obscured AGN model (cstat/dof = 80.68/76; Figure\,\ref{fig:agn}, bottom right). 
The data at higher energy ($>$10\,keV) would be essential to unambiguously determine the true model. The unabsorbed 0.2--10\,keV luminosity of the AGN in this model is $(2.6\pm 0.2)\times 10^{41} \rm erg s^{-1}$, with $94\%$ contribution from the \apec component. 

\begin{figure*}
\centering
\includegraphics[width=0.99\linewidth]{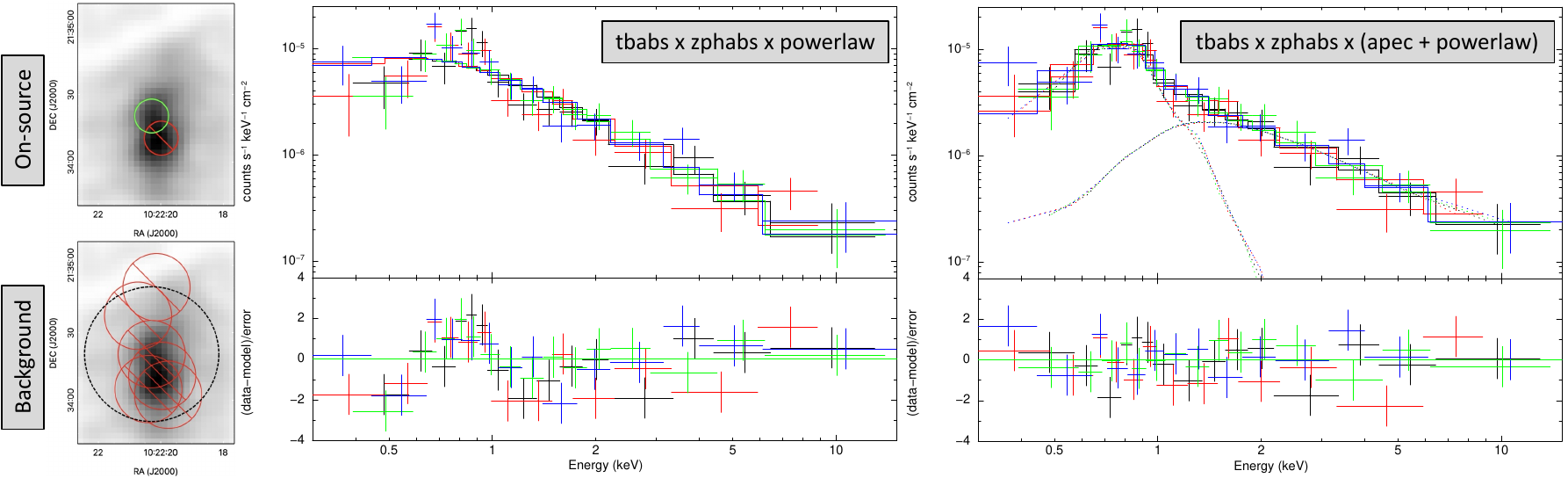}
    \caption{Extraction region (top left) for the new source--1 is shown in green circle, and background region (bottom left) is shown in circle with the regions of red circles masked - plotted over 0.4--12\,keV science image of 0922170301. 0.4--12\,keV spectrum is fitted with an absorbed powerlaw (and discarded; middle), and with an additional absorbed \apec at the redshift of \target (right). ObsIDs 0922170101, 0922170201, 0922170301, and 0922170401 are shown in red, green, black, and blue.}\label{fig:new}
\end{figure*}

{Thus, we have two models that fit the AGN spectra well, and we cannot choose one over the other based on the fit statistic alone. Therefore, we look for other physical constraints to discriminate between them.} The nuclear emission of \target at $12\mu m$ is $\rm log(L^{nuc}_{12\mu m}/erg\, s^{-1}) = 42.60\pm0.06\pm0.43$, based on the empirical scaling relation \citep[hereafter \citetalias{Asmus2014}]{Asmus2014}, and assuming $85\pm10\%$ of the W3 band luminosity is from the nucleus \citepalias{Asmus2020}, with the last source of uncertainty being the observed scatter in their sample. Using the  $\rm L^{nuc}_{12\mu m}-L^{nuc}_{2-10\,keV}$ relation of \citetalias{Asmus2014}, it corresponds to an intrinsic 2--10\,keV nuclear luminosity of $\rm log(L^{nuc}_{2-10\,keV}/erg\, s^{-1}) = 42.30\pm0.06\pm0.34 $, where the last source of uncertainty is the observed scatter in the AGNs in \cite{Asmus2015}. However, our observed X-ray luminosity is $\approx$two orders of magnitude smaller. 
 Thus, if \target satisfies the relation of $\rm L^{nuc}_{12\mu m}-L^{nuc}_{2-10\,keV}$ \citepalias{Asmus2014}, the nucleus of \target has to contribute to $\lessapprox5\%$ of its total W3 band luminosity. It is much smaller than the fraction estimated for the AGN sample (identified based on MIR color, which \target does not satisfy) of \citetalias{Asmus2020}. We can reconcile with this mismatch if the AGN of \target is truly Compton-thick. In that case, the ratio of $12\mu m$ and 2--10\,keV luminosities indicates an absorbing column density of log(N\hin$/\rm cm^{-2}) = 24.2\pm 0.3$ based on the empirical scaling relation of \cite{Asmus2015}. Thus, the Compton-thick hypothesis makes the observed IR and X-ray luminosities consistent with scaling relations, making this our preferred interpretation of the observed X-ray spectrum.

In the spectral models above, we have modeled the soft excess with \texttt{apec}. Because the hot ISM is not systematically brighter in the on-source region than the background region (see contours in Figure\,\ref{fig:contour} for 0.4--0.7\,keV and 0.7--1.0\,keV), the contribution of the diffuse ISM in the soft excess is unlikely to be significant. Therefore, we assign the \apec to the emission from the circumnuclear medium. Assuming a spherical structure, the density of this medium is $(9.5_{-1.6}^{+2.7}) \rm cm^{-3}(r/100 pc)^{-1.5}$. The angular resolution of \xmm limits our ability to distinguish between a $\sim$kpc-long circumnuclear ring that is denser than the rest of the ISM and fosters high star formation, or a $\sim$10--100\,pc circumnuclear disk fueling the AGN.  However, we do not claim the circumnuclear medium to be in equilibrium\footnote{Although, replacing \texttt{apec} with \texttt{NEI} converged the ionization timescale $\tau$ to an equilibrium} or collisionally ionized; it could be photoionized at a lower temperature than the best-fit temperature of \texttt{apec}. The exact physical mechanism of the soft excess in AGNs is generally an open problem, and even in the brightest AGNs, different physical models become statistically indistinguishable \citep[e.g.,][]{Sobolewska2007}. Our main goal was to fit the soft excess with reasonable parameter values that can be used as a prior for more sophisticated, physically motivated models in the future. 

Depending on the spectral model, the unabsorbed 0.3--10\,keV luminosity of the AGN is 5 or 30$\times$ higher than that calculated with an assumed SED in \S\ref{sec:time}. Because XRBs (including ULXs) can also exhibit similar spectral shapes, the \textit{true} luminosity for the ULXs discussed in \S\ref{sec:time} may have been underestimated due to an assumed powerlaw SED. It is further supported by the observed hardness ratio of the ULXs (see Figure\,\ref{fig:rate}, \S\ref{sec:time}), frequently deviating from that expected for the assumed SED. This necessitates the spectral modeling across a broad energy range to achieve an accurate X-ray luminosity estimation of XRBs, which would subsequently affect the shape of the XLF.   

\subsubsection{New source--1}\label{sec:spec_new}

We fit the 0.4--10\,keV spectra of the newly identified source from the four new obsIDs simultaneously with a powerlaw absorbed by the disk of Milky\,Way and the disk of \target (\texttt{tbabs$\times$zphabs$\times$powerlaw}). It does not fit the data well (cstat/dof = 92.02/54) with excess in the data at 0.7--1.1\,keV and deficit in $<0.7$\,keV (Figure\,\ref{fig:new}, middle).

Same as the AGN spectra, we add an absorbed disk blackbody and allow the intrinsic absorption of the powerlaw and blackbody to differ: \texttt{tbabs(zphabs$_1$$\times$diskbb + zphabs$_2$$\times$powerlaw}). It significantly improves the fit (cstat/dof = 41.71/51), requiring the additional component at 100\% confidence (ftest null hypothesis probability 3.8e-7). Tying the two \zph makes the fit worse (cstat/dof =45.19/52), implying the absorbing columns are required to be different by 96\% confidence. However, the normalization of \texttt{diskbb} is egregiously high (ln(norm) = $19.3\pm0.9$), implying an unabsorbed 0.2--10\,keV luminosity of $\rm log(L/erg s^{-1}) = 45.37\pm0.04$, 4--5 orders of magnitude higher than the AGN of \targetn. This indicates accretion at an unphysically large super-Eddington efficiency, for a stellar-/intermediate-mass blackhole. Therefore, we discard the \texttt{diskbb} model on physical grounds. 

Next, we replace \diskbb with \apecn: \texttt{tbabs$\times$zphabs (\apec + powerlaw)}. At the distance to \targetn, the diameter of the spectrum extraction region corresponds to $\approx$ 4.2\,kpc. Thus, the spectra might contain diffuse structure (e.g., a superbubble) in excess of the hot ISM, physically justifying the consideration of \apecn. The revised model fits the data significantly better than the absorbed power-law model (cstat/dof = 41.28/52, Figure\,\ref{fig:new}, right). The unabsorbed 0.2--10\,keV luminosity in this model is $(1.7\pm 0.1)\times 10^{41} \rm erg s^{-1}$, with 87\% contribution from the \apecn. 

\begin{figure*}
\centering
\includegraphics[width=0.99\linewidth]{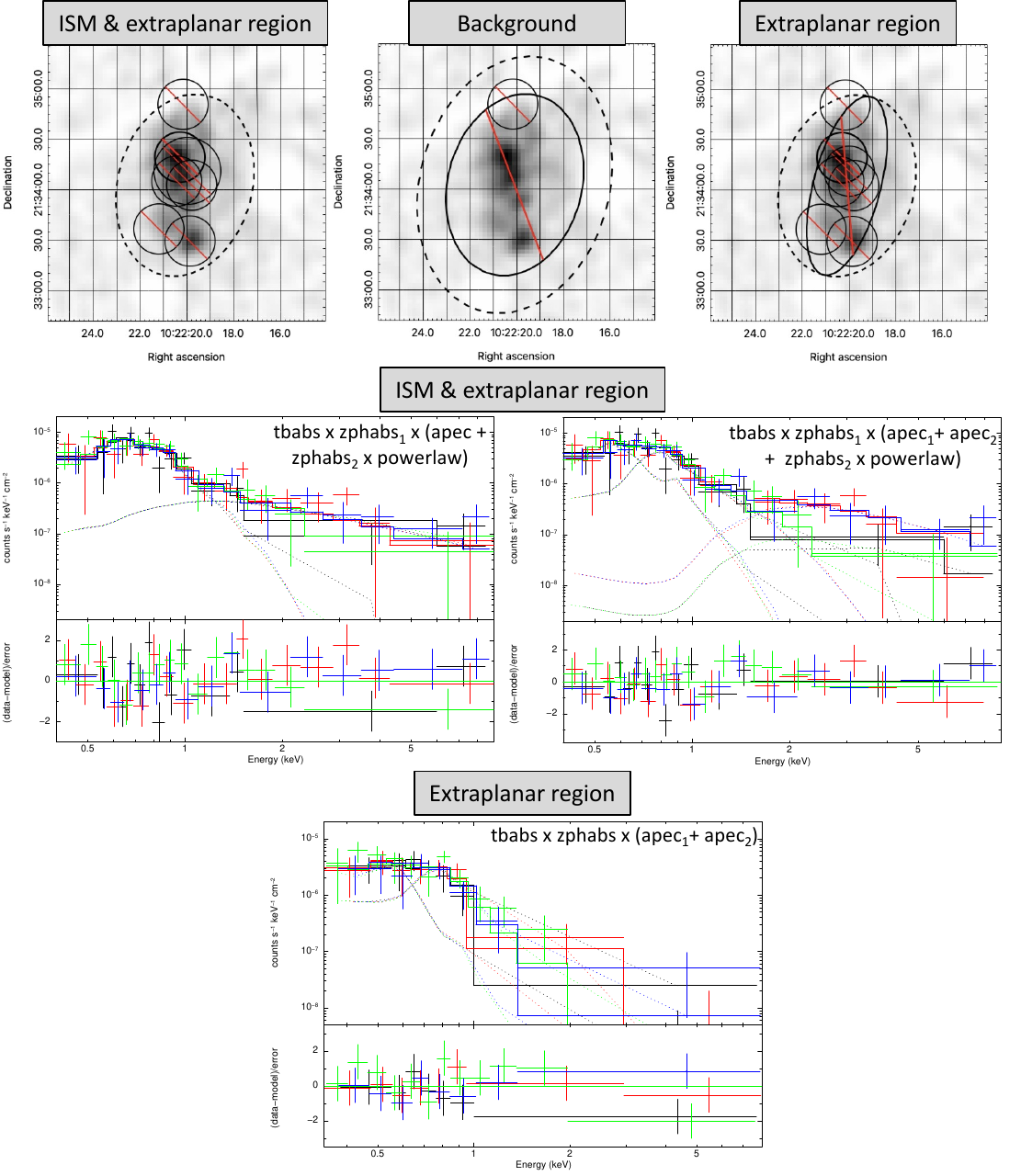}
    \caption{Top: ISM + extraplanar region (left), background region for the hot ISM/extraplanar region (middle), and extraplanar region (right). Because the background for the diffuse medium is defined to be immediately outside the diffuse medium, it provides a conservative estimate of the diffuse medium emission. Middle: Spectra of the ISM+extraplanar region {with a single thermal phase and the powerlaw normalization of all obsIDs tied (left) and the final fit with two thermal phases and powerlaw normalizations tied by part (right)}. Bottom: spectra of the extraplanar region with the best-fitted model. ObsIDs 0922170101, 0922170201, 0922170301, and 0922170401 are shown in red, green, black, and blue. See \S\ref{sec:spec_ism} for more details.} \label{fig:ism}
\end{figure*}

Assuming the \apec component to be volume-filling, the average density of the gas for a spherical structure of 1\,kpc radius is $(2.3_{-0.8}^{+1.4})\times 10^{-1} \rm cm^{-3}(r/1\,kpc)^{-1.5}(Z/Z_\odot)^{-0.5}$. Both in terms of temperature and size-weighted density, it is qualitatively consistent with the previously known superbubbles in our Galaxy (e.g., Local Hot Bubble and Cygnus), M\,31, and NGC\,628 \citep{Liu2017,Bluem2020,Mayya2023,Liu2024}. Thus, it is plausible that we have discovered a superbubble embedded in the ISM of \targetn.   

Note that we cannot rule out the possibility of soft excess associated with the compact object featuring the \pown, while being dominated by the diffuse emission from the superbubble in soft X-ray. Deep, high angular resolution data with sufficient response in soft X-ray ($<1$\,keV) would be essential to infer the actual morphology and physical origin(s) of this new source. 

Same as the AGN, the unabsorbed 0.3--10\,keV luminosity of the ``new" source is $\approx 15\times$ higher than that for an assumed powerlaw SED in \S\ref{sec:time}. This reinstates the importance of broadband spectral modeling for accurately measuring X-ray luminosity of point sources and hence the shape of the XLF. 


\subsubsection{Extended medium}\label{sec:spec_ism}

We fit the ``total diffuse medium" (ISM + extraplanar region) with an \texttt{apec} and the unresolved XRBs with an intrinsically absorbed powerlaw, with both components absorbed by the Galactic disk and by \targetn: \texttt{tbabs$\times$zphabs$_1$$\times$(apec + zphabs$_2$$\times$powerlaw}). The data are not fit well (cstat/dof = 90.98/87), and the deviation of the model from the data in some ObsIDs is primarily visible at $>$1\,keV {(Figure\,\ref{fig:ism}, middle left)}. Therefore, we temporarily untie the powerlaw normalizations of the 4 obsIDs, which improves the fit (cstat/dof = 86.37/84). Then, depending on the similarity among best-fit values, the normalizations of 0922170201 and 0922170301 are tied ($\rm norm_1$ in Table\,\ref{tab:bestfit}); and normalizations of 0922170101 and 0922170401 are tied separately ($\rm norm_2$ in Table\,\ref{tab:bestfit}). The spectra are fit equally well with all powerlaw normalizations varying independently (cstat/dof = 86.37/84) and with this by-part-tied normalization (cstat/dof = 86.80/86). So, we prefer the latter model with 2 fewer variables (hence 2 more degrees of freedom). The adjusted model improves the fit (cstat/dof = 86.80/86), requiring a variable normalization at 95.5\% confidence. 

To test if the diffuse gas originates \textit{in} or \textit{around} the stellar disk of \targetn, we refit the spectra without intrinsic absorption: \texttt{tbabs $\times$ (apec + zphabs$_2$$\times$powerlaw)}. The fit becomes worse (cstat/dof = 93.38/87), requiring the gas to be \textit{in} the disk of \target at 98.8\% confidence. 

Finally, adding another absorbed thermal phase, \texttt{tbabs $\times$ zphabs$_1$ $\times$ (apec$_1$ + apec$_2$ + zphabs$_2$$\times$ powerlaw)}, provides a better fit than a single thermal phase (cstat/dof=78.61/84), requiring the additional \apec component at 98.4\% confidence (Figure\,\ref{fig:ism}, middle right). The excess emission fit by the second \apec cannot be fit by tuning the chemical composition of the first \apec or replacing the first \apec with a non-equilibrium model. 

The $\rm norm_2$ is larger than $\rm norm_1$ by a factor of $4.1_{-0.7}^{+5.9}$, implying significant variability of the unresolved sources in the disk. The total count rate of resolved point sources discussed in \S\ref{sec:time} does not vary in the same fashion; thus, the spilled emission from those sources is unlikely to cause the variability seen in these spectra. The corresponding unabsorbed 0.2--10\,keV luminosity of the powerlaw is $\rm log(L/erg s^{-1}) = 39.39_{-0.74}^{+0.26}$ and $\rm log(L/erg s^{-1}) = 40.02_{-0.10}^{+0.08}$. Because these observations were performed within 9 days, the variable luminosity of unresolved sources reinstates the importance of considering a variable XLF discussed in \S\ref{sec:time}. 

The unabsorbed 0.2--10\,keV luminosity of \apecn$_1$ and \apecn$_2$ are $\rm log(L/erg s^{-1}) = 40.76\pm0.06$ and $\rm log(L/erg s^{-1}) = 39.72\pm0.05$. The combined 0.5--2\,keV luminosity of these two components is $\approx3-4\times$ higher than nearby galaxies of similar SFR \citep{Strickland2004a,Owen2009,Mineo2012b}, putting it above the observed scatter of 0.34\,dex around the best-fitted $\rm L^{gas}_{X}-SFR$ relation in the literature. Assuming $\rm 10^{51}erg$ of mechanical energy per supernova (SN) and using the prescriptions of \cite{Heckman1990,Mannucci2005}, the total energy output from type Ia and core-collapse supernovae for the stellar mass and SFR of \target is $\rm \dot{E}_{SN} = 25.4 \pm 3.0 \times 10^{41} erg\,s^{-1}\Big(\frac{E_{SN}}{10^{51}erg}\Big)$. The resulting conversion efficiency of the SN mechanical energy into the ISM thermal energy is $\eta = \frac{\rm L^{gas}_{X}}{\rm \dot{E}_{SN}} = 4.8\pm0.8$\%. This suggests that the remaining energy might have been expelled outside the disk and extraplanar region. 

\begin{table*}[]
    \caption{Best fit parameters for the AGN, the new source--1, and the diffuse medium in and around the stellar disk}
    \begin{tabular}{cccccc}
    \hline
    \hline
     \multicolumn{6}{c}{AGN} \\
\multicolumn{3}{c}{Obscured} & \multicolumn{3}{c}{Compton-thick} \\  
    \hline
    Components & Parameters & Values & Components & Parameters & Values \\
    \hline
     \zphn$_1$ &   N(\hin) [cm$^{-2}$] & $5.7_{-0.9}^{+1.0}\times 10^{21}$ & \zph & N(\hin) [cm$^{-2}$] &  $1.3\pm 0.1\times 10^{22}$\\
     \diskbb &    T [keV] & 0.21$\pm$0.02 & \apec & T [keV] & $0.20\pm0.02$\\
     \diskbb &    norm & $1.6_{-0.9}^{+2.1}$ & \apec & norm & $3.8_{-1.3}^{+2.2}\times 10^{-4}$\\
     & & & \pex & rel$\rm_{refl}$ & $-0.79\mp$0.07\\
     \zphn$_2$ &  N(\hin) [cm$^{-2}$] & $3.4_{-0.8}^{+0.9}\times 10^{22}$ & \pex & E$_c$ [keV] & $26.4_{-9.0}^{+28.2}$\\
     \pow &    $\Gamma$ & $0.9\pm0.2$ & \pex & $\Gamma$ & $2.2\pm0.1$ \\
     \pow &    norm & $2.9_{-0.8}^{+1.2}\times 10^{-6}$ & \pex & norm & $7.1\pm0.8\times 10^{-4}$\\ 
    \hline
     \multicolumn{6}{c}{New source--1} \\
    \hline
    & & Components & Parameters & Values & \\
    \hline
     & & \zph & N(\hin) [cm$^{-2}$] &  $9.2_{-2.3}^{+1.4}\times 10^{21}$\\
     & & \apec & T [keV] & $0.21_{-0.02}^{+0.04}$ \\
     & & \apec & ln(norm) & $-8.4_{-1.2}^{+0.8}$ \\
     & & \pow & $\Gamma$ & $1.6\pm0.1$ \\
     & & \pow & norm & $ 7.3\pm1.1 \times 10^{-6}$\\ 
    \hline
      \multicolumn{3}{c}{ISM + extraplanar region} & \multicolumn{3}{c}{Extraplanar region} \\
    \hline
    Components & Parameters & Values & Components & Parameters & Values \\
    \hline
     \zphn$_1$ &   N(\hin) [cm$^{-2}$] & $5.9\pm0.2\times 10^{21}$ & \zph & N(\hin) [cm$^{-2}$] &  $3.0_{-2.0}^{+2.2}\times 10^{20}$\\
     \apecn$_1$ & T [keV] & $0.14_{-0.07}^{+0.02}$ & \apecn$_1$ & T [keV] & $0.16_{-0.04}^{+0.06}$ \\
     \apecn$_1$ & norm & $1.0_{-0.4}^{+7.0} \times 10^{-4}$ & \apecn$_1$ & norm & $1.8_{-0.5}^{+2.1} \times 10^{-6}$\\
     \apecn$_2$ & T [keV] & $0.49_{-0.17}^{+0.09}$ & \apecn$_2$ & T [keV] & $0.61\pm0.10$\\
     \apecn$_2$ & norm & $5.8_{-1.6}^{+8.5} \times 10^{-6}$ & \apecn$_2$ & norm & $0.8_{-0.2}^{+0.1} \times 10^{-6}$ \\
     \zphn$_2$ &  N(\hin) [cm$^{-2}$] & $2.8_{-0.8}^{+1.1}\times 10^{22}$ \\
     \pow & $\Gamma$ & $2.0\pm0.5$ & \\
     \pow & norm$_1$ & $1.0_{-0.8}^{+1.2}\times 10^{-6}$ \\
     \pow & norm$_2$ & $4.1_{-2.1}^{+3.4}\times 10^{-6}$ \\
    \hline  
    \hline
    \end{tabular}
    \label{tab:bestfit}
\end{table*}

Next, we fit the extraplanar region with \texttt{tbabs $\times$ zphabs $\times$ apec}. It provides a reasonably good fit (cstat/dof=51.69/53), but a two-\apec model,  \texttt{tbabs $\times$ zphabs $\times$ (apec$_1$ + apec$_2$)}, provides a better fit (cstat/dof=44.40/51), requiring the additional component at 97.9\% confidence (Figure\,\ref{fig:ism}, bottom). The unabsorbed 0.2--10\,keV luminosity of apec$_1$ and apec$_2$ are $\rm log(L/erg s^{-1}) = 38.88_{-0.06}^{+0.05}$ and $\rm log(L/erg s^{-1}) = 39.04_{-0.10}^{+0.08}$.

The temperature of the hot (\apecn$_2$) phase is $3.5_{-1.3}^{+1.9}$ times higher in the ISM+extraplanar region and $3.8\pm1.3$ times higher in the extraplanar region than the warm-hot (\texttt{apec$_1$}) phase (see Table\,\ref{tab:bestfit}). However, the normalization of \texttt{apec$_1$} decreases by two orders of magnitude outside the disk, while for \texttt{apec$_2$} the normalization decreases by one order of magnitude. This is naturally reflected by the luminosity, with the hot phase being brighter than the warm-hot phase in the extraplanar region. This indicates a flatter surface brightness profile of the hot phase than the warm-hot phase. A more extended, hotter phase indicates that the average temperature of the diffuse highly ionized medium might increase with height from the stellar disk (i.e., a ``temperature inversion"). By considering the measured temperature as a density-squared weighted average of the underlying temperature distribution (${\rm T_{obs}} = \int n^2{\rm T}dl/\int n^2 dl$), we calculate the average temperature of the warm-hot and hot phases. The average temperature of the ISM and the extraplanar region turns out to be $\sim$0.16\,keV and $\sim$0.30\,keV, respectively, revealing a ``temperature inversion" by a factor of 2. This is consistent with the soft band contours being more extended than the super-soft band contours (\S\ref{sec:imaging}; Figure\,\ref{fig:contour}).

The specific supernovae rate of \targetn, $\approx 88$ Myr$^{-1}$kpc$^{-2}$ \citep{Strickland2004a,Henley2010}, is larger than the critical surface rate for superbubble blowout of $\rm 25-40\,Myr^{-1}kpc^{-2}$ \citep{Maclow1988}. Thus, the ``temperature inversion" could potentially happen due to starburst-driven winds lifting the hot gas outside the disk. The D-array radio (21-cm) continuum data of \target (which is more sensitive to diffuse emission) is more extended than the C-array data \citep{Irwin1999}, implying the nonthermal heating by cosmic-ray electrons could also be responsible for the ``temperature inversion". Determining the relative contribution of stellar/AGN feedback and the hot virialized CGM of \target in the ``temperature inversion" is beyond the scope of this paper; it would be part of a follow-up investigation (Pan, Das \textit{et al.}, in prep.). 

Because \target is edge-on, it is reasonable to assume that the spatial extent of the diffuse ISM along the line-of-sight is similar to its spatial extent in the sky plane along the stellar disk. Assuming that the diffuse medium is volume filling ($f_V=1$) and at solar metallicity, the average density of the warm-hot and hot phases in the ISM+extraplanar region are ${\rm 5.0_{-0.8}^{+16.9}\times 10^{-3}cm^{-3}}$ and ${\rm 1.2_{-0.2}^{+0.9}\times 10^{-3}cm^{-3}}$, respectively. In the extraplanar region these densities reduce to ${\rm 7.5_{-1.0}^{+4.4}\times 10^{-4}cm^{-3}}$ and ${\rm 5.0_{-0.6}^{+0.3}\times 10^{-4}cm^{-3}}$. This, together with the measured temperatures, shows that the X-ray surface brightness decreases from the ISM to the extraplanar region due to the lower density and not lower temperature, contrary to the common assumption in imaging-only analyses. The thermal pressure, $\rm P/k_B$, of the warm-hot and hot phases in the ISM+extraplanar region are $8.1_{-4.3}^{+27.5}\times 10^3 \rm cm^{-3}K$ and $6.8_{-2.6}^{+5.3}\times 10^3 \rm cm^{-3}K$, implying these phases are in pressure equilibrium. However, in the extraplanar region, the pressure of the hot phase, $3.5_{-0.7}^{+0.6}\times 10^3 \rm cm^{-3}K$, is $2.5_{-1.3}^{+4.2}$ times larger than that of the warm-hot phase, $1.4_{-0.4}^{+1.0}\times 10^3 \rm cm^{-3}K$. Thus, the additional $2.1_{-1.2}^{+0.7}\times 10^3 \rm cm^{-3}K$ pressure from cooler phases and/or non-thermal sources might sustain the extraplanar region in pressure equilibrium, or the excess pressure of the hot phase might lead to a snowplowed warm-hot phase surrounding the hot phase. The volume filling factor and metallicity could differ across phases as well as with heights from the stellar disk, so one can obtain \textit{true} densities and thermal pressures using the scaling relations of $n \propto {\rm (Z/Z_\odot)^{-0.5}} f_V^{-0.5}$ for an appropriate physical model of the multiphase highly ionized ISM and extraplanar region. 

\section{Summary and future directions}\label{sec:summary}
We present deep \xmm observations of a nearby star-forming luminous infrared galaxy \target and discuss the properties of its nuclear and disk-wide point sources as well as diffuse gas through images, light curves, and spectra in 0.4--12\,keV. It is an interesting target for the following reasons: 
\begin{enumerate}
    \item We confirm the existence of a low-luminosity AGN based on its unambiguous detection in 5--12\,keV. We explore two options in the spectral analysis: a heavily obscured AGN with soft ($<1$\,keV) X-ray emission from the accretion disk, or a Compton-thick AGN with its reflected component and circumnuclear medium visible in the \xmm spectra. Distinguishing between them requires further investigation in $>10$\,keV. 
    \item We study the light curves of the six ULXs previously reported in the \chandra data. We find diverse variability timescales in the broadband luminosity and a wide range of hardness ratios (which are also variable) of the ULXs. The combined emission from unresolved point sources also indicates day-scale variability in the spectral analyses. This provides us with crucial insights into the short-scale time dependence and stochasticity of the X-ray luminosity function of galaxies.  
    \item We identify two distinct bright regions in the stellar disk that are detected in the archival (although not reported before) and the new \xmm data, but not in the \chandra data. One of them is a supersoft source emitting exclusively in $<$1\,keV, and mostly in $<$0.7\,keV. The other one is the brightest soft X-ray emitter in the disk and reveals a unique and puzzling spectral shape. In addition to a compact object with an absorbed powerlaw spectrum, a superbubble in \target could explain the soft excess. A blackbody associated with the compact object is also considered for the soft excess and ruled out due to the resulting unphysically high luminosity. 
    \item Spectral analysis of the diffuse gas in and around the stellar disk reveals two phases at $\sim$0.15\,keV and $\sim$0.55\,keV. The hot phase is an order-of-magnitude fainter than the warm-hot phase in the disk. But the former is more extended beyond the disk, indicating a ``temperature inversion". This is also consistent with the different shapes of the narrowband (0.4--0.7\,keV and 0.7--1.0\,keV) intensity contours. 
    \item The detected and unresolved point sources (except the AGN) contribute to $\approx 42$\% of the 0.2--10\,keV unabsorbed intrinsic luminosity, making the diffuse component the primary contributor to the X-ray emission, which we attribute to its high SFR.    
\end{enumerate}

Obtaining multiple snapshots of the same galaxy with X-ray instruments featuring high angular resolution and broad energy coverage (from 0.1\,keV to $>$10\,keV) is essential to put detailed and simultaneous constraint on the spatial, spectral, and temporal variability of (resolved and unresolved) X-ray binaries, AGN, and the hot gas including the diffuse ISM, supperbubbles, and the extraplanar region. 

\section*{acknowledgments}
S.D. acknowledges \xmm AO22 award 80NSSC23K1513. S.D. is grateful for the support provided by NASA through the NASA Hubble Fellowship grant HST-HF2-51551.001-A awarded by the Space Telescope Science Institute, which is operated by the Association of Universities for Research in Astronomy, Inc., under NASA contract NAS5-26555. S.M. is grateful for the NASA ADAP grant 80NSSC22K1121. Y.K. acknowledges support from UNAM-PAPIIT grant IN102023. 

This work is based on observations obtained with \xmmn, an ESA science mission with instruments and contributions directly funded by ESA Member States and NASA. {This paper employs a list of \chandra datasets, obtained by the \chandra X-ray Observatory, contained in the \chandra Data Collection \dataset[DOI: 10398]{https://doi.org/10.25574/cdc.497}}, and published previously in \citetalias{Luangtip2015}. We thank Wasutep Luangtip for providing us with the precise sky location of the ULXs reported in \citetalias{Luangtip2015}. We thank the anonymous referee for the constructive suggestions. This research has made use of NASA's Astrophysics Data System Bibliographic Services. 



\facilities{\xmmn,\chandran}

\software{{\texttt{AstroPy} \citep{Astropy2013,Astropy2018,Astropy2022}}, \texttt{DS9} \citep{Joye2003},  \texttt{HeaSoft} \citep{Drake2005}, \texttt{Jupyter} \citep{jupyter2016}, \texttt{Matplotlib} \citep{Hunter2007}, \texttt{NumPy} \citep{numpy2020}, \texttt{\href{https://github.com/astropy/pyregion}{pyregion}}, \texttt{SAS} \citep{Snowden2004}, \texttt{SciPy} \citep{scipy2022}, \texttt{XSPEC} \citep{Arnaud1999}}
    
\bibliographystyle{aasjournal}

\end{document}